\documentclass[lettersize,journal]{IEEEtran}
\usepackage{amsmath,amsfonts}
\usepackage{algorithmic}
\usepackage{algorithm}
\usepackage{array}
\usepackage[caption=false,font=normalsize,labelfont=sf,textfont=sf]{subfig}
\usepackage{textcomp}
\usepackage{stfloats}
\usepackage{url}
\usepackage{verbatim}
\usepackage{graphicx}
\usepackage{cite}
\usepackage{svg}
\usepackage{booktabs}
\usepackage{multirow}    
\usepackage{threeparttable} 
\begin{document}

\title{Metapopulation Graph Neural Networks: \\ Deep Metapopulation Epidemic Modeling with Human Mobility$\dagger$}

\author{Qi Cao, Renhe Jiang\IEEEauthorrefmark{1}, Chuang Yang, Zipei Fan, Xuan Song, Ryosuke Shibasaki
	
\IEEEcompsocitemizethanks{
\IEEEcompsocthanksitem \IEEEauthorrefmark{1} Corresponding to Renhe Jiang.
\IEEEcompsocthanksitem Q. Cao, R. Jiang, C. Yang, Z. Fan, X. Song, R. Shibasaki are with Center for Spatial Information Science, The University of Tokyo; X. Song is also with SUSTech-UTokyo Joint Research Center on Super Smart City, Department of Computer Science and Engineering, Southern University of Science and Technology; E-mail: \{caoqi,jiangrh,chuang.yang\}@csis.u-tokyo.ac.jp; fanzipei@iis.u-tokyo.ac.jp; \{songxuan,shiba\}@csis.u-tokyo.ac.jp
\IEEEcompsocthanksitem \textbf{$\dagger$ This is the extended version of the ECMLPKDD2022 paper ~\cite{cao2023mepognn}. The main incremental changes include: a mobility generation method and the experiments to test its effectiveness, the detailed descriptions of data processing, the visualization and analysis of data, the limitation discussion based on extra test data.}
}
}

\maketitle

\begin{abstract}
Epidemic prediction is a fundamental task for epidemic control and prevention. Many mechanistic models and deep learning models are built for this task. However, most mechanistic models have difficulty estimating the time/region-varying epidemiological parameters, while most deep learning models lack the guidance of epidemiological domain knowledge and interpretability of prediction results. In this study, we propose a novel hybrid model called MepoGNN for multi-step multi-region epidemic forecasting by incorporating Graph Neural Networks (GNNs) and graph learning mechanisms into Metapopulation SIR model. Our model can not only predict the number of confirmed cases but also explicitly learn the epidemiological parameters and the underlying epidemic propagation graph from heterogeneous data in an end-to-end manner. The multi-source epidemic-related data and mobility data of Japan are collected and processed to form the dataset for experiments. The experimental results demonstrate our model outperforms the existing mechanistic models and deep learning models by a large margin. Furthermore, the analysis on the learned parameters illustrate the high reliability and interpretability of our model and helps better understanding of epidemic spread. In addition, a mobility generation method is presented to address the issue of unavailable mobility data, and the experimental results demonstrate effectiveness of the generated mobility data as an input to our model.
\end{abstract}

\begin{IEEEkeywords}
Epidemic forecasting, Hybrid model, Metapopulation epidemic model, Graph learning, GNNs, COVID-19
\end{IEEEkeywords}

\section{Introduction}
The coronavirus disease 2019 (COVID-19) pandemic has caused around 500 million confirmed cases and more than 6 million deaths in the global, and it is still ongoing. Due to this circumstance, epidemic forecasting has been a key research topic again as it can guide the policymakers to develop effective interventions and allocate the limited medical resources. Many mechanistic models and deep learning models have been built for the epidemic prediction task. In particular, human mobility is seen as one of the most important factors to understand and forecast the epidemic propagation among different regions. 

\begin{figure}[h]
	\centering
    \includegraphics[width=1.0\linewidth]{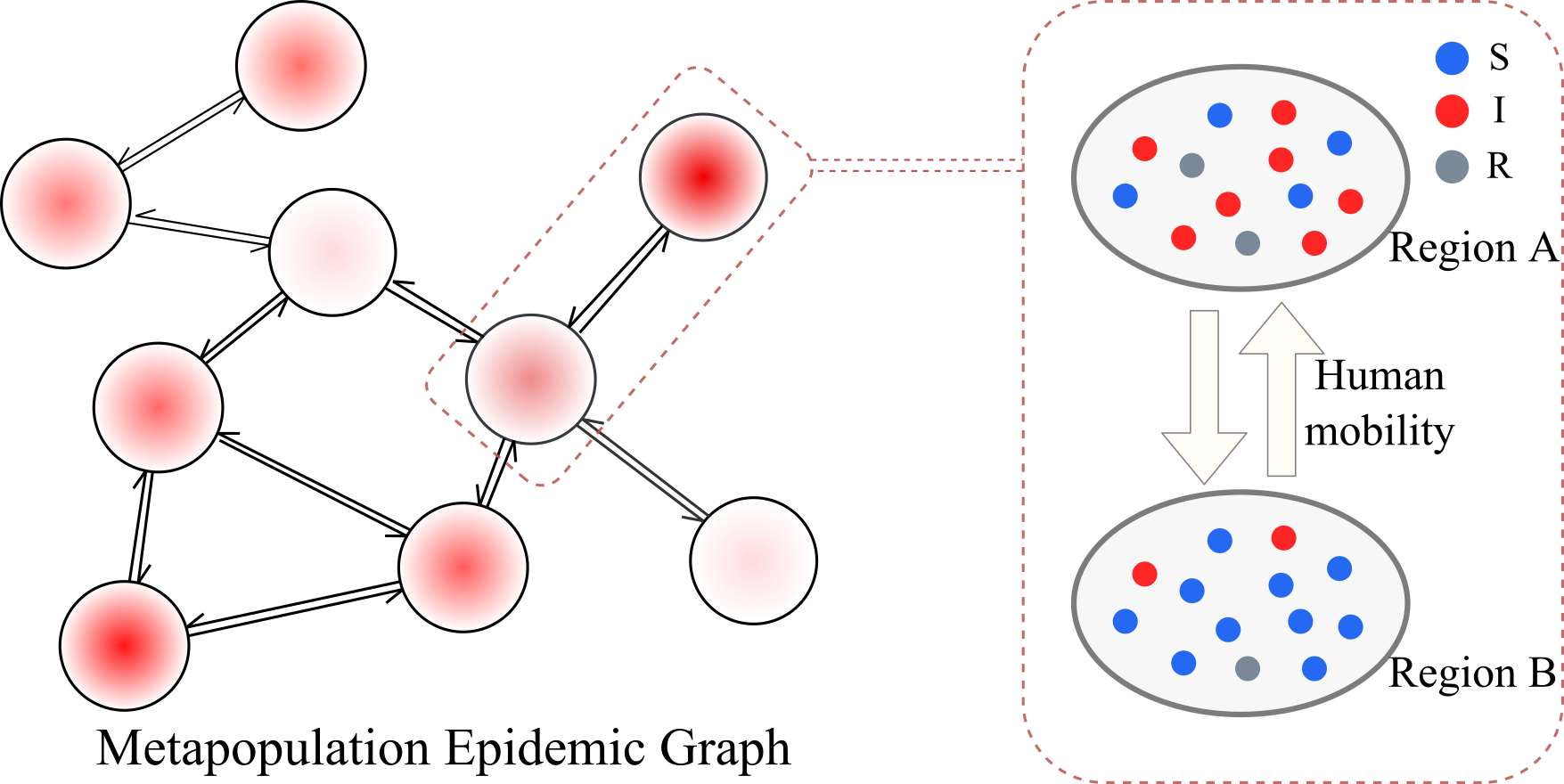}
	\caption{Metapopulation epidemic propagation among regions \protect \cite{ref_1}.}
	\label{fig:intro}
\end{figure}

In this study, we employ metapopulation SIR model~\cite{ref_1,ref_2} as the base model for our task, which extends the most fundamental compartmental model (i.e., SIR \cite{ref_11}) in epidemiology with metapopulation epidemic propagation. As illustrated in Fig. \ref{fig:intro}, it divides the total population under the epidemic into several sub-populations (e.g., by regions). Each sub-population consists of three compartments, $S$ (susceptible individuals), $I$ (infectious individuals), $R$ (removed individuals, including deaths and recovery cases), and the human mobility between sub-populations is modeled as a directed graph. Thus, it can well model the epidemic propagation among regions. The metapopulation epidemic models have achieved great success in modeling and analyzing the propagation of epidemic diseases, such as SARS, H1N1, and Malaria~\cite{ref_3,ref_4,ref_5}.

However, it is always a non-trivial task to build a metapopulation epidemic model, especially for new emerging epidemics such as the COVID-19 due to the following reasons. First, the epidemiological parameters in metapopulation model keep varying from region to region and time to time. As we all know, the Coronavirus keeps evolving, and the transmissibility and mortality of the variants (e.g., Alpha, Delta, and Omicron) are significantly different. Besides, the intervention policies and the human movements also vary over different periods and regions. Second, due to the mixed factors mentioned above, the epidemic propagation effects via human mobility in metapopulation epidemic model are also difficult to be obtained or estimated. In the case of prefecture-level prediction in Japan, we need to collect the large-scale human mobility data of the entire Japan and obtain the amount of human movements between each pair of prefectures. Then how to accurately infer the underlying disease propagation network becomes another intractable task. Third, besides the daily infection data, external features such as date information (e.g., $dayofweek$) and daily movement change patterns should also be involved.

To tackle these challenges, we incorporate deep learning into metapopulation SIR model to form a novel hybrid epidemic model. Specifically, we first learn the time/region-varying epidemiological parameters from multiple data features through a spatio-temporal module, which consists of Temporal Convolutional Networks (TCN) and Graph Convolutional Networks (GCN). Next, we design two types of graph learning module to automatically approximate the underlying epidemic propagation graph based on the countrywide human mobility data. Furthermore, we let the learned latent graph be shared by the spatio-temporal module and the metapopulation SIR module, which further enhances the model interpretability and reliability. Previous deep learning methods~\cite{ref_6,ref_7,ref_8,ref_9,ref_10} simply treat the epidemic forecasting as time-series prediction task or spatio-temporal prediction task, which can only output the predicted number of infections in a pure black-box manner. Recent study~\cite{ref_29} involves the classical epidemic modeling into deep neural networks. However, it does not explicitly consider the epidemic propagation among regions via metapopulation modeling like our model, which largely limits the model interpretability for multi-region epidemic forecasting. \emph{To the best of our knowledge, our work is the first hybrid model that couples metapopulation epidemic model with spatio-temporal graph neural networks}. In summary, our work has the following contributions:
\begin{itemize}
    \item We propose a novel hybrid model along with two types of graph learning module for multi-step multi-region epidemic prediction by mixing metapopulation epidemic model and spatio-temporal graph convolution networks.
	\item Our model can explicitly learn the time/region-varying epidemiological parameters as well as the latent epidemic propagation among regions from the heterogeneous inputs like infection related data, human mobility data, and meta information in a completely end-to-end manner. 
	\item We collect and process the big human GPS trajectory data and other COVID-19 related data that covers the 47 prefectures of Japan from 2020/04/01 to 2021/09/21 for countrywide epidemic forecasting. 
	\item We conduct comprehensive experiments to evaluate the performance of our model on epidemic forecasting task of the 47 prefectures in Japan by comparing with three classes of baseline models. The results illustrate the superior forecasting performance of our model, especially for unprecedented surge of cases. Furthermore, we also use the case studies to demonstrate the high interpretability of our model.
	\item We present a mobility generation method with minimal data requirement to handle the situation which mobility data is unavailable. The experimental results show the effectiveness of generated mobility data as the initialization of adaptive graph learning in our model. 
\end{itemize}

\section{Related Work}\label{sec:relatedworks}
In this section, we first generally review the two major classes of models for epidemic forecasting and then introduce the research directly related to our model with more details. 

\subsection{Epidemic Forecasting Models}
The models for epidemic simulation and forecasting can be divided into two types: \textit{mechanistic approaches} and \textit{deep learning approaches}.

\textit{Mechanistic approaches} are built based on the domain knowledge of epidemiology which employ pre-defined physical rules to model infectious diseases' transmission dynamics, mainly \textit{classical compartmental models}~\cite{ref_11,ref_12}, \textit{metapopulation models}~\cite{ref_2,ref_13,ref_14,ref_15} and \textit{agent-based model}~\cite{ref_16,ref_17,ref_18}.
The classical compartmental models simulate the
spread of infectious diseases in a homogeneous population which are unable to model epidemic spread between regions.
The metapopulation models assume the heterogeneity of sub-populations and use the human mobility pattern between regions to model the spread of the epidemic~\cite{ref_1,ref_2}. 
The agent-based models directly use the individual-level movement pattern~\cite{ref_16,ref_17} or trajectories~\cite{ref_18} to emulate the contagion process. 
Our work is related to the metapopulation model which is most suitable for multi-region epidemic forecasting task. 
To implement epidemic modeling, it needs to be calibrated first using historical observations and use the optimized or manually modified parameters to make prediction. 
These efforts are hardly applicable for multi-step forecasting tasks. The parameters calibration process needs high computational complexity, especially when facing huge parameter state space~\cite{ref_16,ref_13}. Moreover, in most mechanistic models, epidemiological parameters keep fixed during forecasting. The variation of parameters through time is not considered which leads to the problem of cumulative error on multi-step prediction. 

\textit{Deep learning approaches} have shown excellent performance in the modeling and forecasting on time series prediction tasks. 
As a typical time series, several research efforts utilizing deep learning techniques, such as LSTM~\cite{ref_6,ref_8}, have been conducted for epidemic forecasting over a single region~\cite{ref_6,ref_8,ref_19,ref_20}. Nevertheless, the epidemic propagation is often spatially dependent, i.e., co-evolving over regions. Thus, 
treating epidemic forecasting as a multivariate time-series prediction task, performing collaborative forecasting over multiple geographical units should be a more reasonable choice.
For such tasks, a key challenge is to model the complex and implicit spatio-temporal dependencies among the observations, on which much evidence shows that GNN can perform very well for modeling the inter-series relationships. A series of state-of-the-art solutions based on GNN have been proposed for multivariate time-series prediction tasks, such as STGCN~\cite{ref_21}, DCRNN~\cite{ref_22}, GraphWaveNet~\cite{ref_23}, ColaGNN~\cite{ref_9}, and CovidGNN~\cite{ref_10}. 
In particular, ColaGNN~\cite{ref_9} and CovidGNN~\cite{ref_10} were explicitly designed for the epidemic prediction. However, these works ignore the domain knowledge of epidemiology and are hard to interpret from the epidemiological perspective. STAN~\cite{ref_19} incorporates epidemiological constraints into deep learning models, but it can only predict infections of a single region. CausalGNN~\cite{ref_29} embeds single-patched SIRD model into GNN for multi-region epidemic forecasting.

Overall, we distinguish our work from existing ones in the following ways: 
Compared with the mechanistic models, MepoGNN adopts an end-to-end framework that can predict the dynamic change of epidemiological parameters and use predicted parameters to produce multi-region and multi-step prediction; 
Compared with the deep learning models for the multi-region prediction task, MepoGNN incorporates the domain knowledge of epidemiology and enhances the interpretability by combining spatio-temporal deep learning model with the metapopulation model; 
Furthermore, MepoGNN can output the prediction of infections through the metapopulation epidemic model and learn the interpretable epidemiological parameters and the latent graph of epidemic propagation simultaneously. 

\subsection{SIR Model and Metapopulation SIR Model}
\begin{figure}
	\centering
	\includegraphics[width=0.8\linewidth]{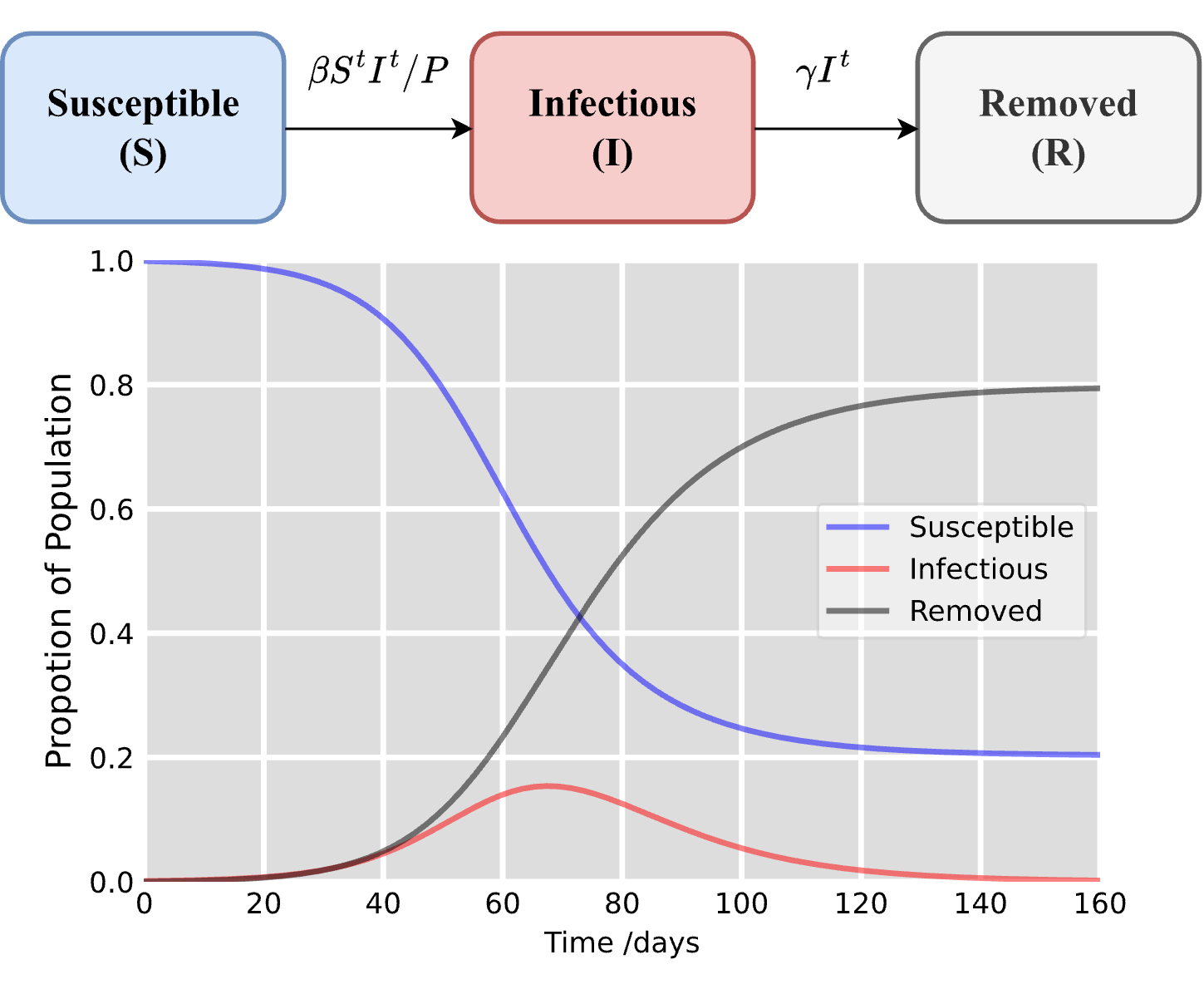}
        \caption{Epidemic spread in classical SIR model.}
	\label{fig:classicalSIR}
\end{figure}

After the general review of epidemic forecasting models, we introduce the models closely related to our research with more details. The compartmental models are widely used technique for modeling the spread of infectious diseases. SIR model \cite{ref_11} is one of the most representative classical compartmental models (most other compartmental models, like SEIR model, SIRD model and SIRV model can be seen as variants of SIR model). SIR model divides population into three compartments, including $S^t$ for number of susceptible individuals, $I^t$ for number of infectious individuals, $R^t$ for the number of recovered or deceased individuals at time $t$. As shown in Fig. \ref{fig:classicalSIR}, susceptible individuals can become infectious individuals when they contact with infectious individuals and get infected by the infectious disease, and infectious individuals can become removed individuals when they recover or die from the infectious disease. And SIR model uses $\beta$ as the infection rate and $\gamma$ as the removal rate. $S$, $I$, $R$ can be updated by following equations:
\begin{equation}\label{classicalSIR}
    \begin{aligned}
    \frac{dS^{t+1}}{dt} & =  -\frac{\beta S^t I^t}{P} \\
    \frac{dI^{t+1}}{dt} & =  \frac{\beta S^t I^t}{P} - \gamma I^t\\
    \frac{dR^{t+1}}{dt} & =  \gamma I^t
    \end{aligned}
\end{equation}
where $P$ represents total population size which is usually assumed as a constant number ($P = S^t + I^t + R^t$). 

\begin{figure}
	\centering
    \includegraphics[width=1.0\linewidth]{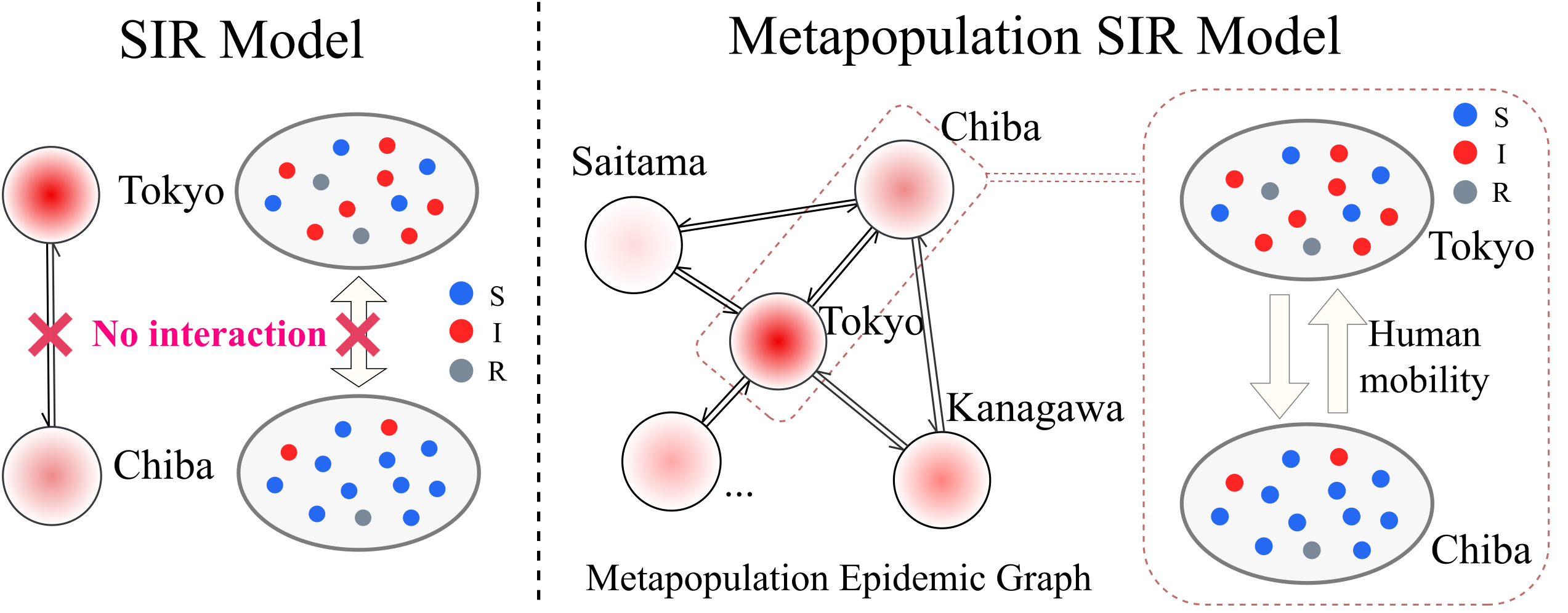}
	\caption{Differences between SIR model and Metapopulation SIR model.}
	\label{fig:meta_sir_compare}
\end{figure}

When SIR model is used for epidemic modeling, the typical method is as follows:
\begin{enumerate}
    \item Input historical epidemic into Eq. \ref{classicalSIR} and optimize parameters (i.e., $\beta$ and $\gamma$) to fit the epidemic curve.
    \item Use optimized parameters to update $S$, $I$, $R$ iteratively to model the epidemic spread in future.
\end{enumerate}

SIR model can only model the epidemic spread for a homogeneous population, which ignores the epidemic propagation between sub-populations. However, in the real world, the sub-populations usually have heterogeneous epidemic situations and interact with each other (e.g., epidemic propagation among sub-populations). Metapopulation SIR model \cite{ref_2} fills this gap by assuming the heterogeneity of sub-populations and using human mobility to model the propagation between sub-populations. We demonstrate the difference between SIR model and Metapopulation SIR model by an example of Tokyo and Chiba in Fig. \ref{fig:meta_sir_compare}. Metapopulation SIR model consists of three compartments for each sub-population: $S_{n}^t$ for number of susceptible individuals, $I_{n}^t$ for number of infectious individuals, $R_{n}^t$ for the number of recovered or deceased individuals of sub-population $n$ at time $t$. $P_{n}$ represents the size of sub-population $n$ which is assumed to be a constant number, where $P_{n}=S_{n}^t+I_{n}^t+R_{n}^t$. $\beta$ is the rate of infection\footnote{$\beta$ in metapopulation SIR model is not completely equivalent to $\beta$ in SIR model, but we use the same symbol in this work for simplicity.}, and $\gamma$ is the rate of recovery and mortality. Furthermore, it uses $h_{nm}$ to represent the epidemic propagation from sub-population $n$ to $m$. The original metapopulation SIR model in \cite{ref_2} can be expressed as follows:
\begin{equation}\label{metasir}
    \begin{aligned}
    \frac{dS_{n}^{t+1}}{dt} & =  -\beta\cdot S_{n}^t\mathop{\sum_{m=1}^{N}(\frac{h_{mn}}{P_{m}}+\frac{h_{nm}}{P_{n}})I_{m}^t} \\
    \frac{dI_{n}^{t+1}}{dt} & = \beta\cdot S_{n}^t\mathop{\sum_{m=1}^{N}(\frac{h_{mn}}{P_{m}}+\frac{h_{nm}}{P_{n}})I_{m}^t}-\gamma\cdot I_{n}^t \\
    \frac{dR_{n}^{t+1}}{dt} & =  \gamma\cdot I_{n}^t
    \end{aligned}
\end{equation}
The parameter $h_{nm}$ can form an epidemic propagation graph as shown in Fig. \ref{fig:meta_sir_compare}. In \cite{ref_2}, the task is formed as a network inference problem, and optimization for the objective function with some regularization is used to solve this problem. 

\begin{figure}[h]
\centering
\includegraphics[width=0.9\linewidth]{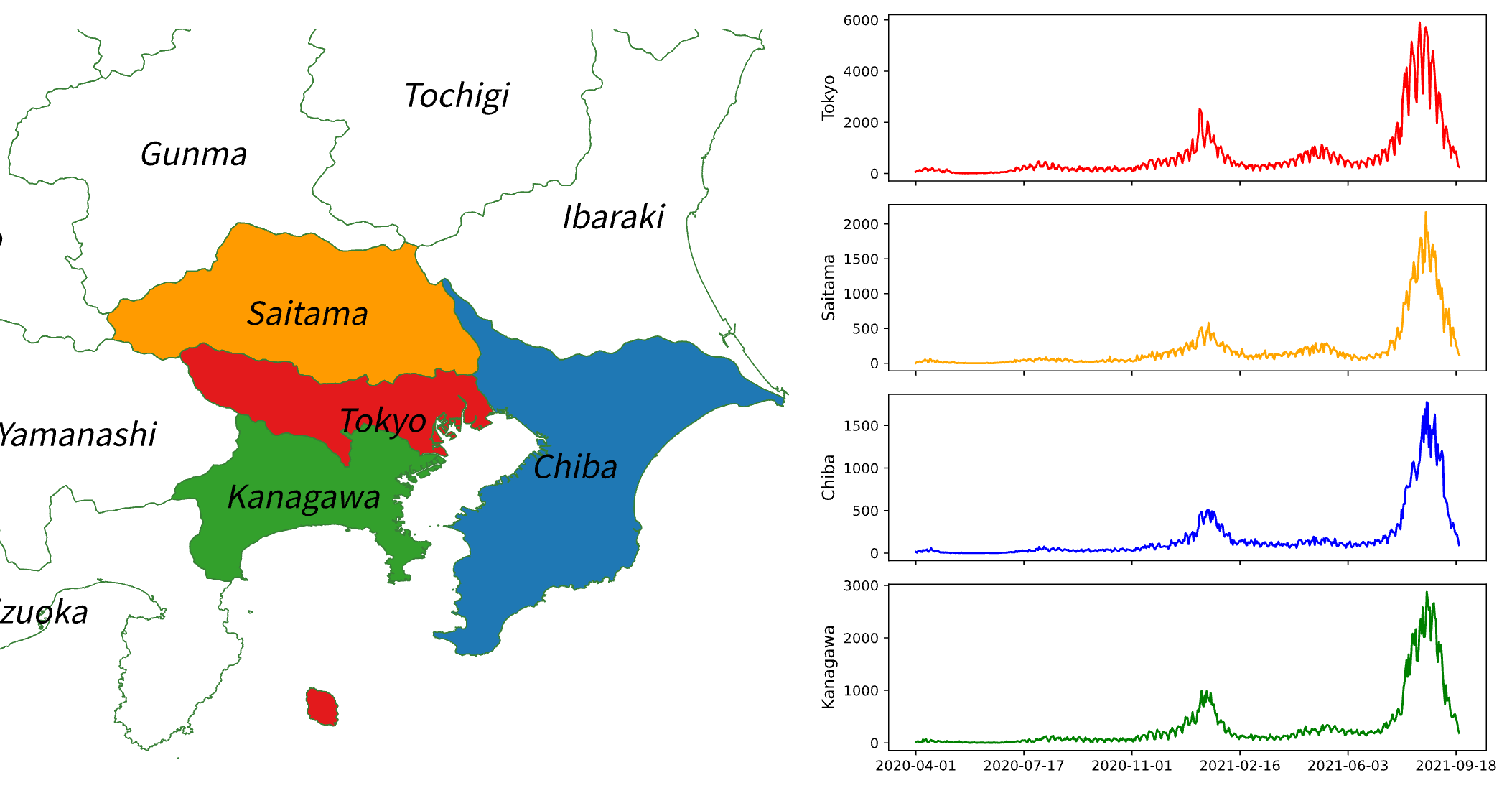}
\caption{Illustration of multi-regional epidemic forecasting.}
\label{fig:multiregion}
\end{figure}

\begin{figure}[h]
\centering
\includegraphics[width=1.0\linewidth]{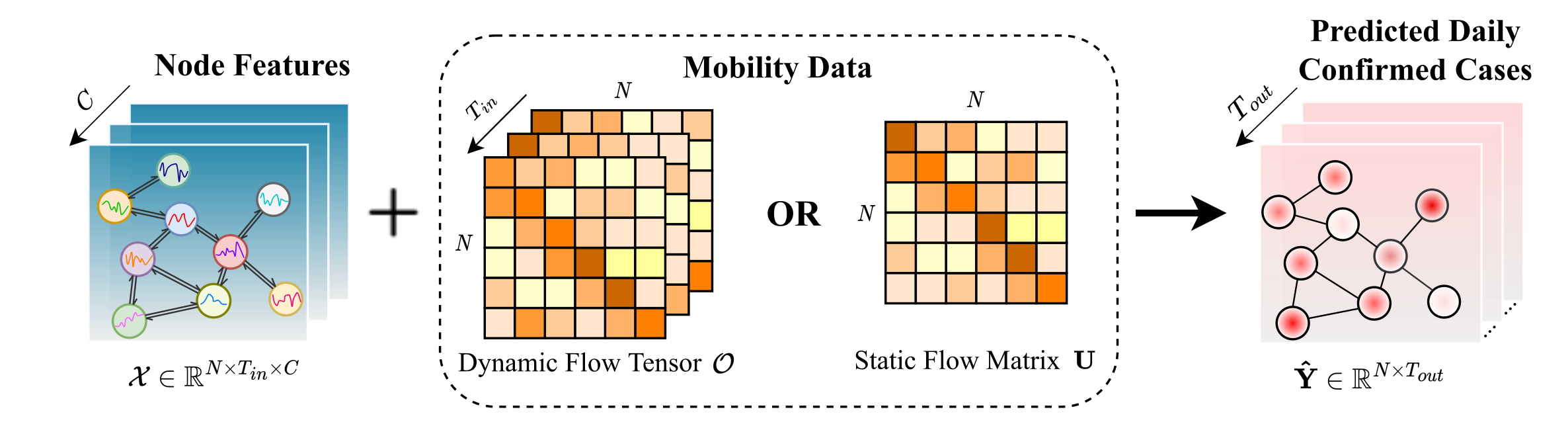}
\caption{Illustration of problem definition (model input and output).}
\label{fig:problemdefinition}
\end{figure}

\section{Problem Formulation}\label{sec:problemdefinition}
In this study, we focus on forecasting the number of daily confirmed cases for multi-region and multi-step simultaneously by using epidemic related data and human mobility data. 

For a single region, the historical daily confirmed cases from timestep $t-T_{in}+1$ to $t$ can be represented as $\mathbf x^{t-(T_{in}-1):t} \in \mathbb{R}^{T_{in}}$. Then, the  historical daily confirmed cases of $N$ regions as illustrated in Fig. \ref{fig:multiregion} can be denoted as $\mathbf X^{t-(T_{in}-1):t} = \{\mathbf x^{t-(T_{in}-1):t}_1, \mathbf x^{t-(T_{in}-1):t}_2, ..., \mathbf x^{t-(T_{in}-1):t}_N\} \in \mathbb{R}^{N \times T_{in}}$. Besides the historical observations, we also incorporate the external factors to form a multi-channel input as $\mathcal{X}^{t-(T_{in}-1):t} =\{\mathbf{X}^{t-(T_{in}-1):t}_1, \mathbf{X}^{t-(T_{in}-1):t}_2, ...,  \mathbf{X}^{t-(T_{in}-1):t}_C\} \in \mathbb{R}^{N \times T_{in} \times C}$. Details of the input features will be introduced in Section~\ref{sec:data}. Additionally, human mobility between regions (static flow data $\mathbf{U} \in \mathbb{R}^{N \times N}$ or dynamic flow data $\mathcal{O}^{t-(T_{in}-1):t} \in \mathbb{R}^{N \times N \times T_{in}}$) is used as another type of input. Details of processing mobility flow data will be also introduced in Section~ \ref{sec:data}.  

Thus, in this study, we aim to predict the daily confirmed cases of $N$ regions in next $T_{out}$ timesteps $\mathbf{Y}^{t+1:t+T_{out}} \in \mathbb{R}^{N \times T_{out}}$ that takes human mobility between regions into consideration. As demonstrated by Fig. \ref{fig:problemdefinition}, our problem can be formulated as follows:
\begin{equation}
\{\mathcal{X}^{t-(T_{in}-1):t}, \mathbf{U}\} \xrightarrow{ \quad {f}(\cdot) \quad} \mathbf{Y}^{t+1:t+T_{out}}
\end{equation}
\begin{equation}
\{\mathcal{X}^{t-(T_{in}-1):t}, \mathcal{O}^{t-(T_{in}-1):t}\} \xrightarrow{ \quad {f}(\cdot) \quad} \mathbf{Y}^{t+1:t+T_{out}}
\end{equation}

\section{Methodology}\label{sec:method}
\begin{figure*}[b!]
    \centering
     \includegraphics[width=0.9\textwidth]{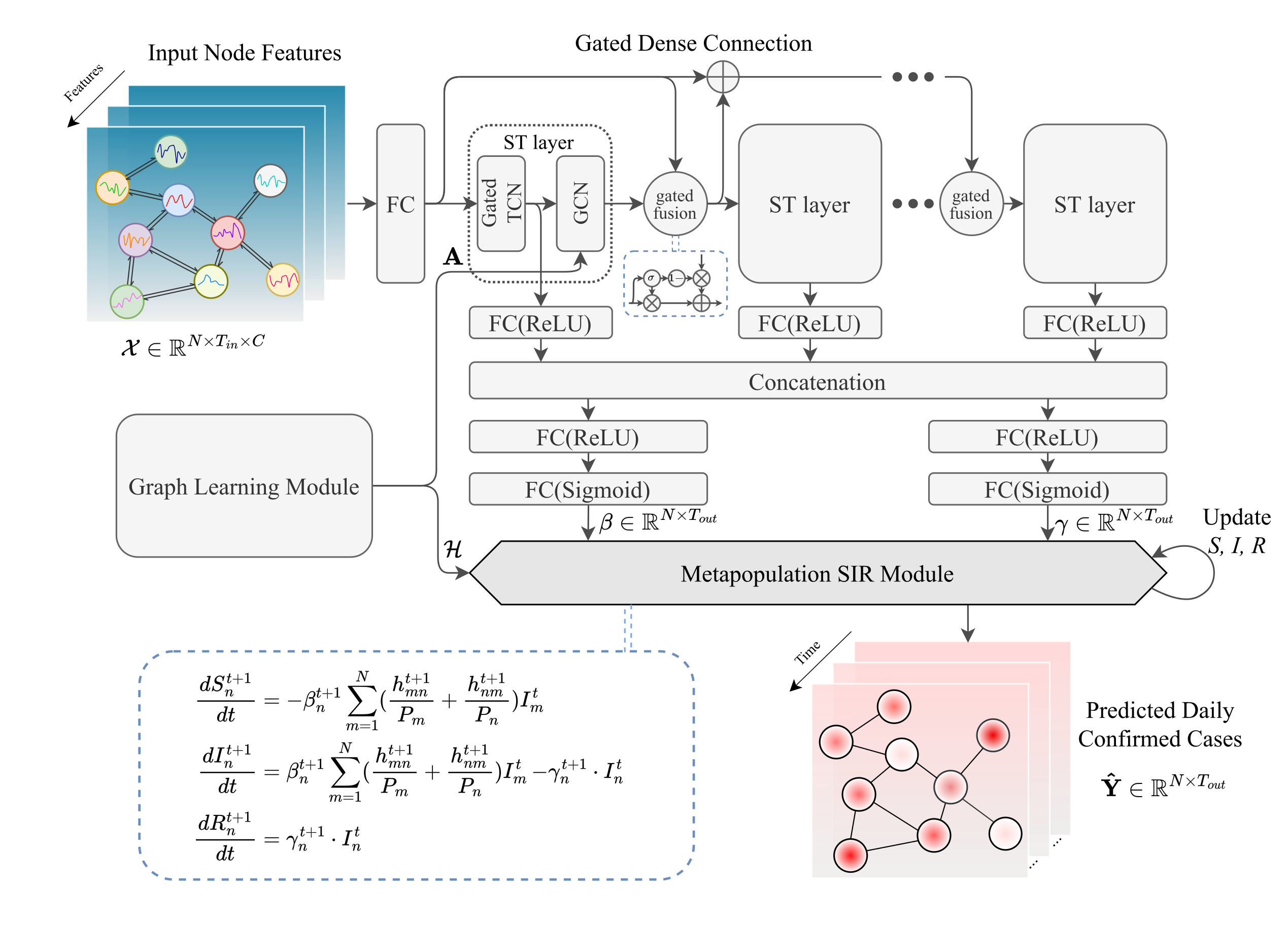}
     \caption{Proposed Epidemic Metapopulation Graph Neural Networks (MepoGNN) for spatio-temporal epidemic prediction.}
     \label{fig:mepognn}
\end{figure*}

In this section, we present Epidemic Metapopulation Graph Neural Networks (MepoGNN), demonstrated in Fig. \ref{fig:mepognn}, for spatio-temporal epidemic prediction. MepoGNN consists of three major components: metapopulation SIR module, spatio-temporal module and graph learning module. These three components tightly cooperate with each other. Graph learning module learns the mobility intensity between regions as a graph and output it to spatio-temporal module and metapopulation SIR module. Spatio-temporal module captures the spatio-temporal dependency to predict the sequences of parameters for metapopulation SIR module. Then, metapopulation SIR module takes the learned graph and the predicted parameters to produce the multi-step prediction of daily confirmed cases. 

\begin{figure*}[h]
    \centering
    \includegraphics[width=0.9\textwidth]{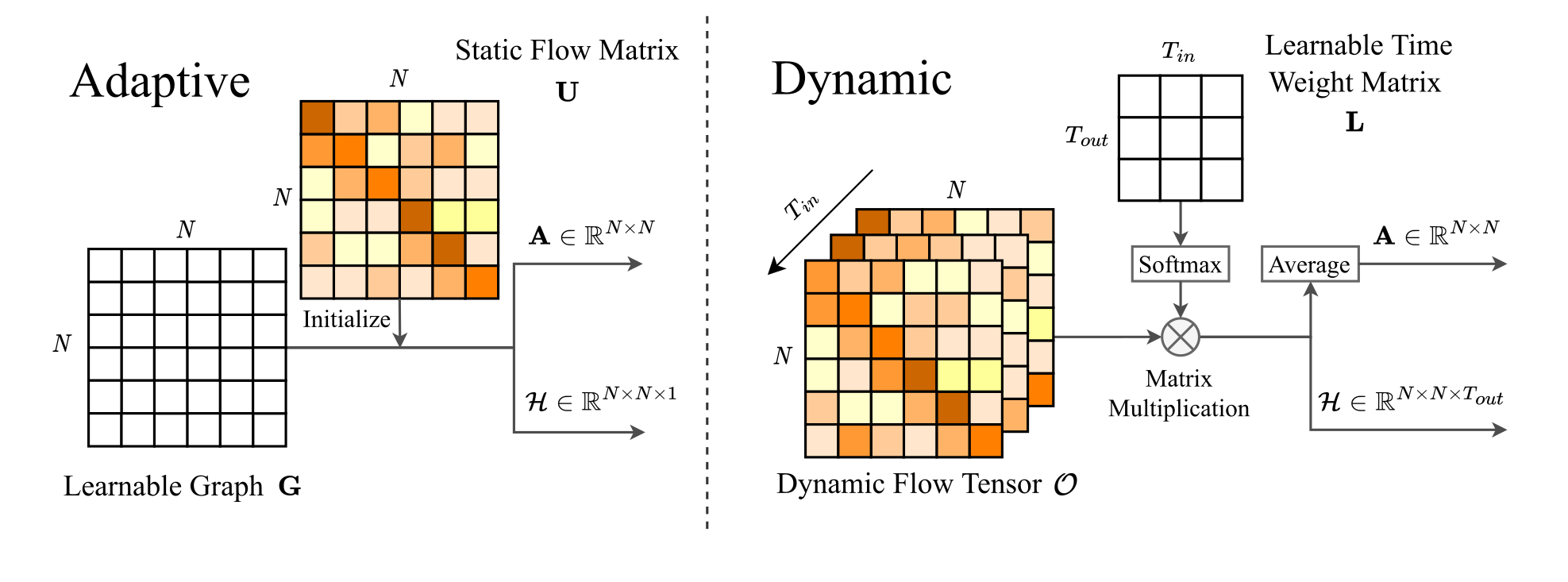}
    \caption{Two types of graph learning: Adaptive and Dynamic.}
    \label{fig:graphlearning}
\end{figure*}

\subsection{Metapopulation SIR Module}
SIR model is one of the most fundamental compartmental models in epidemiology, used for modeling the epidemic spread \cite{ref_11}. However, it can only model the epidemic spread for a homogeneous population, which ignores the epidemic propagation between sub-populations. Metapopulation SIR model \cite{ref_2} fills this gap by assuming the heterogeneity of sub-populations and using human mobility to model the propagation between sub-populations. 

In this study, we model population of each region as sub-population in metapopulation SIR model. So, the $h_{nm}$ in Eq. \ref{metasir} can be represented by human mobility between regions. Because of different characteristics of regions, policy changes with time and so on, there exists spatio-temporal heterogeneity in epidemic spread. In our model, $\beta$, $\gamma$ and $h_{nm}$ are assumed to vary over time and regions. In addition, to prevent $\beta$ to be extremely small and make it be in a relatively stable magnitude, $S_{n}^t$ is omitted from the equations. Thus, we extend the original metapopulation SIR in Eq. \ref{metasir} as follows:
\begin{equation}\label{eq:metaSIR}
    \begin{aligned}
    \frac{dS_{n}^{t+1}}{dt} & =-\beta_{n}^{t+1}\mathop{\sum_{m=1}^{N}(\frac{h_{mn}^{t+1}}{P_{m}}+\frac{h_{nm}^{t+1}}{P_{n}})I_{m}^t} \\
    \frac{dI_{n}^{t+1}}{dt} & =  \beta_{n}^{t+1}\mathop{\sum_{m=1}^{N}(\frac{h_{mn}^{t+1}}{P_{m}}+\frac{h_{nm}^{t+1}}{P_{n}})I_{m}^t}-\gamma_{n}^{t+1}\cdot I_{n}^t\\
    \frac{dR_{n}^{t+1}}{dt} & =  \gamma_{n}^{t+1}\cdot I_{n}^t
    \end{aligned}
\end{equation}

With predicted $\beta_{n}^{t+1}$, $\gamma_{n}^{t+1}$ and $\mathcal{H}^{t+1}$ (the epidemic propagation matrix formed by $\{h_{nm}^{t+1} | n,m\in\{1,2,...,N\}\}$), $S$, $I$, $R$ can be updated iteratively:
\begin{equation}\label{updateSIR}
[S_{n}^t, I_{n}^t, R_{n}^t] \xrightarrow[\beta_{n}^{t+1}, \gamma_{n}^{t+1}, \mathcal{H}^{t+1}]{Eq. (\ref{eq:metaSIR})} [S_{n}^{t+1}, I_{n}^{t+1}, R_{n}^{t+1}]
\end{equation}
The final prediction output of daily confirmed cases can be formed as:
\begin{equation}\label{finaloutput}
    \begin{aligned}
    {\hat{y}_{n}^{t+1}} &= \beta_{n}^{t+1}\mathop{\sum_{m=1}^{N}(\frac{h_{mn}^{t+1}}{P_{m}}+\frac{h_{nm}^{t+1}}{P_{n}})I_{m}^{t}} \\
    \mathbf{\hat{Y}} &= \begin{bmatrix}
\hat{y}_{1}^{t+1} & \dots & \hat{y}_{1}^{t+T_{out}}\\
\vdots  & \ddots & \vdots\\
\hat{y}_{n}^{t+1} & \dots & \hat{y}_{n}^{t+T_{out}}
\end{bmatrix}_{N \times T_{out}}
    \end{aligned}
\end{equation}

\subsection{Spatio-Temporal Module for Epidemiological Parameters}
Spatio-temporal module takes the node input features $\mathcal{X}\in \mathbb{R}^{N \times T_{in} \times C}$ and the weighted adjacency matrix $\textbf{A}\in \mathbb{R}^{N \times N}$ as input and output the predicted parameters $\beta \in \mathbb{R}^{N \times T_{out}}$ and $\gamma \in \mathbb{R}^{N \times T_{out}}$. We use the spatio-temporal layer (ST layer) combining Gated TCN and GCN (same as in GraphWaveNet \cite{ref_23}) to capture the spatio-temporal dependency. Gated TCN \cite{ref_24} is used to capture temporal dependency: 
\begin{equation}
    \mathcal{Q}_{l} = g({\Theta_{l1}} \star \mathcal{Z}_{l} + \mathbf{b}_{l1}) \odot  \sigma({\Theta_{l2}} \star \mathcal{Z}_{l} + \mathbf{b}_{l2}) 
\end{equation}
where $\mathcal{Z}_{l}$ is input of $l$-th layer, ${\Theta_{l1}}$ and ${\Theta_{l2}}$ are temporal convolution kernels, $\mathbf{b}_{l1}$ and $\mathbf{b}_{l2}$ are biases, $g(\cdot)$ is tanh activation function for output, $\sigma(\cdot)$ is sigmoid function to form the gate, $\star$ is convolution, $\odot$ is element-wise product. Next, we model the regions and the interactions between regions as a graph and use diffusion graph convolution \cite{ref_22,ref_23} to capture the spatial dependency:
\begin{equation}\label{eq: AtoP}
\begin{aligned}
    \textbf{P}_{f} &= \textbf{A}/rowsum(\textbf{A}) \\
    \textbf{P}_{b} &= \textbf{A}^{\textbf{T}}/rowsum(\textbf{A}^{\textbf{T}}) 
\end{aligned}
\end{equation}
\begin{equation}
        \tilde{\mathcal{Z}}_{l} = \sum_{k=0}^K \textbf{P}^{k}_{f}\mathcal{Q}_{l}\mathbf{W}_{lk1} + \textbf{P}^{k}_{b}\mathcal{Q}_{l}\mathbf{W}_{lk2}
\end{equation}
where $\mathbf{A}\in \mathbb{R}^{N \times N}$ is weighted adjacency matrix, $\textbf{P}_{f}$ is forward transition matrix, $\textbf{P}_{b}$ is backward transition matrix, $\tilde{\mathcal{Z}}_{l}$ is output of $l$-th layer.

Multiple ST layers can be stacked to capture the spatio-temporal dependency in different scales. We use a gated dense connection to bridge different ST layers. It can extract important information from previous ST layers and pass it to following layer:
\begin{equation}
    \mathcal{D}_{l} = 
    \begin{cases}
        \mathcal{X}, &\text{if}\ l = 1,\\
        \mathcal{D}_{l-1} + \mathcal{Z}_{l}, & \text{otherwise}. 
    \end{cases}
\end{equation}
\begin{align}
    \mathcal{Z}_{l+1} = \begin{cases}
        \mathcal{X}, & \text{if}\ l = 0,\\
        \tilde{\mathcal{Z}}_{l} \odot \sigma(\tilde{\mathcal{Z}}_{l})+ \mathcal{D}_{l} \odot (1 - \sigma(\tilde{\mathcal{Z}}_{l})), & \text{otherwise}. 
        \end{cases}
\end{align}
where $\mathcal{D}_{l}$ stores the information from previous layers. Then, we concatenate the output from different layers through skip connections to fuse the information of different scales. Finally, the parameters $\beta\in \mathbb{R}^{N \times T_{out}}$ and $\gamma\in \mathbb{R}^{N \times T_{out}}$ are produced through two fully connected layers, respectively. 

\subsection{Graph Learning Module for Epidemic Propagation}
There are two different graphs used in metapopulation SIR module and spatio-temporal module, respectively. Unlike the trivial method which input two fixed graphs to each module separately, we make two modules share a single learnable graph. With the shared learnable graph, the spatial dependency used in spatio-temporal module would be consistent with epidemic propagation in metapopulation SIR module which can improve the interpretability of our model. Furthermore, the parameters of graph learning module can be updated by gradients from both spatio-temporal module and metapopulation SIR module which make learned graph more realistic.

As shown in Fig. {\ref{fig:graphlearning}}, there are two types of graph learning module to deal with different input data. The first type is adaptive graph learning module which takes the static flow data (e.g., commuter survey data) as input. Unlike the graph learning methods in some deep learning models \cite{ref_23,ref_26,ref_27}, it is difficult to random initialization for graph learning module in our model. Because the metapopulation SIR module requires the input $\mathcal{H}^t$ to be a graph with values which can reflect the real heterogeneous mobility intensity between each pair of regions (e.g., the number of trips, the number of commuters) in an appropriate magnitude (to make the parameters easier to learn) rather than a normalized (e.g., through row normalization) weighted adjacency graph. 
Intuitively, we initialize an adaptive graph $\textbf{G}$ with static flow matrix $\mathbf{U}$ and make it learnable through training. Then, the adaptive graph can be output to spatio-temporal module (Eq. \ref{eq: AtoP}) as $\textbf{A}\in \mathbb{R}^{N \times N}$ and to metapopulation SIR module (Eq. \ref{eq:metaSIR}) as $\mathcal H \in \mathbb{R}^{N \times N \times 1}$ (which means we use same $h_{nm}$ for all timesteps). 

The second type is dynamic graph learning module which takes the dynamic OD flow tensor as input. Although the OD flow and epidemic spread status are both dynamic, but they are not necessarily one-to-one temporally corresponding. Considering the delayed effect, influence of mobility on epidemic spread can be seen as a weighted average of the given past values ($T_{in}$ days). So, we initialize a learnable time weight matrix $\textbf{L}\in \mathbb{R}^{T_{out} \times T_{in}}$ and normalize it as $\mathbf{\tilde{L}}$ through a softmax function. The normalized time weight matrix can map the historical dynamic flow $\mathcal O^{t-(T_{in}-1):t} \in \mathbb{R}^{N \times N \times T_{in}}$ to its influence on future epidemic spread. The output of $\mathcal H^{{t+1}:{t+T_{out}}} \in \mathbb{R}^{N \times N \times T_{out}}$ and $\textbf{A}\in \mathbb{R}^{N \times N}$ can be calculated as follows:
\begin{equation}
    \mathbf{\tilde{L}} = Softmax_{:,j}(\mathbf{L})
\end{equation}
\begin{equation}
    \mathcal H^{{t+1}:{t+T_{out}}} = \mathbf{\tilde{L}} \mathcal O^{t-(T_{in}-1):t}, \quad \mathbf{A} = \frac{\sum_{i=1}^{T_{out}}\mathcal H^{t+i}}{T_{out}}
\end{equation}

\noindent\textbf{Why propose two types of graph learning?} Dynamic graph learning module can illustrate the dynamic change of epidemic propagation. But it requires dynamic flow data which is not available in most cases. To improve the applicability of our model, we propose adaptive graph learning module to address this problem. With two types of graph learning module, our model can handle different situations of data availability in the best way possible. 

\subsection{Mobility Generation for Initialization}
As mentioned above, we proposed two types of graph learning module to handle different situations of data availability. However, it is required to access static flow data even for adaptive graph learning module. However, in some worse situations of data availability, the static flow data is difficult to access or even not exist. So, how to deal with this problem is a key to improve the applicability of our MepoGNN model. 

In this section, we try to generate the mobility between each pair of regions using as simple data as possible. In the assumption, there are two major factors determining the mobility intensity between each pair of regions: 

(1) Populations of origin and destination: Population of each region determines the trip generation from each region and trip attraction to each region. For example, the metropolitan regions (e.g., Tokyo) have more capability to generate the trip from itself to other regions and also attract more visitors from other regions.

(2) Distance between origin and destination: Distance between origin and destination is also a key factor determining the mobility distribution. Each region tends to generate more trips to nearer regions and attract more visitors from nearer regions. For example, Aichi Prefecture and Saitama Prefecture have similar populations, but there are more trips from Tokyo to Saitama than from Tokyo to Aichi. 

The relationship between the mobility intensity between from region $n$ to $m$ and the above-mentioned two factors can be simply approximated as following equation:
\begin{equation}
    mob_{nm} \propto \frac{P_n P_m}{(dist_{nm})^d + \epsilon}
\end{equation}
where $P_n$ and $P_m$ is populations of region $n$ and $m$, respectively; $dist_{nm}$ is distance between region $n$ and $m$; $d$ is a parameter to control the power of distance decay; $\epsilon$ is a constant number to prevent dividing by zero when computing intra-region mobility (i.e., $n$ = $m$).

In MepoGNN model, we use mobility between regions to estimate the epidemic propagation between regions. So, the absolute mobility intensity (e.g., real trip numbers between each pair of regions) is not required in our study. The relative mobility intensity can be approximated by following equation:

\begin{equation}\label{relativemob}
    \widetilde{mob}_{nm} = \alpha \frac{P_n P_m}{(dist_{nm})^d+ \epsilon}
\end{equation}
where $\alpha$ is a parameter to control $\widetilde{mob}_{nm}$ in an appropriate magnitude. 

The relative mobility among $N$ regions $\{\widetilde{mob}_{nm} | n,m\in\{1,2,...,N\}\}$ can form a relative mobility intensity matrix $\mathbf{\widetilde M} \in \mathbb{R}^{N \times N}$ which can be used to initialize the learnable graph of adaptive graph learning module. 
Although this simple assumption might not provide a precise estimation of mobility intensity between regions, it is sufficient as initialization for adaptive graph learning module since the learnable graph can evolve during training. Furthermore, the aim of our method is to generate relative mobility intensity using as simple data as possible. To best of our knowledge, population of each region and distance between regions are available in most cases and can be seen as the most simple data requirement for mobility generation. This minimal data requirement ensures better applicability of our method. 

\begin{figure}[t]
    \centering
    \includegraphics[width=1.0\linewidth]{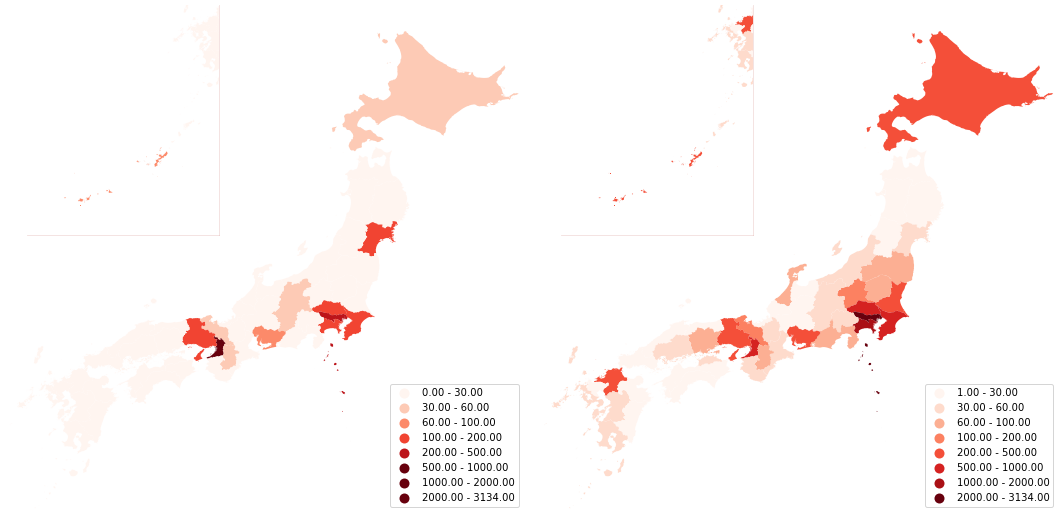}
    \caption{Spatial distribution of daily confirmed cases of Japan on 2021/04/01 (left) and 2021/08/01 (right).}
    \label{fig:spatial_distribution}
\end{figure}

\begin{figure}[t]
    \centering
    \includegraphics[width=1.0\linewidth]{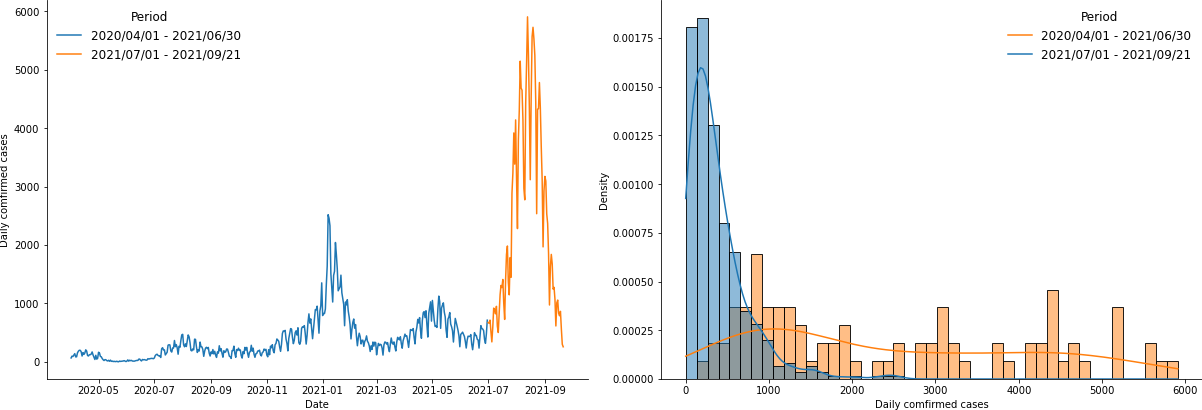}
    \caption{Daily confirmed cases of Tokyo (left) and distribution gap between two periods of time (right).}
    \label{fig:tokyo_distribution}
\end{figure}

\begin{figure}[t]
    \centering
    \includegraphics[width=1.0\linewidth]{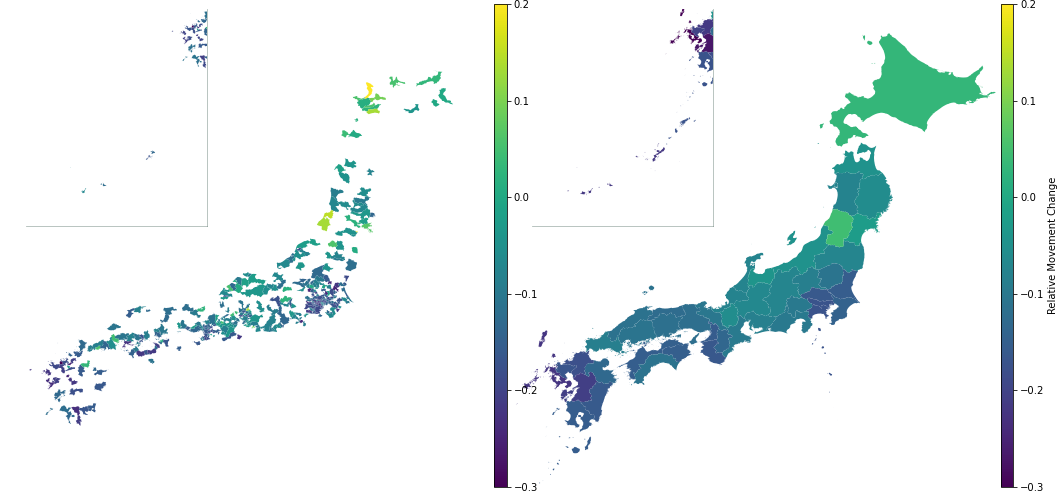}
    \caption{Original movement change data (left) and prefecture-level movement change data (right) through population weighted averaging on 2021/08/01.}
    \label{fig:movement_popweight}
\end{figure}

\begin{figure}[t] 
    \centering
    \includegraphics[width=1.0\linewidth]{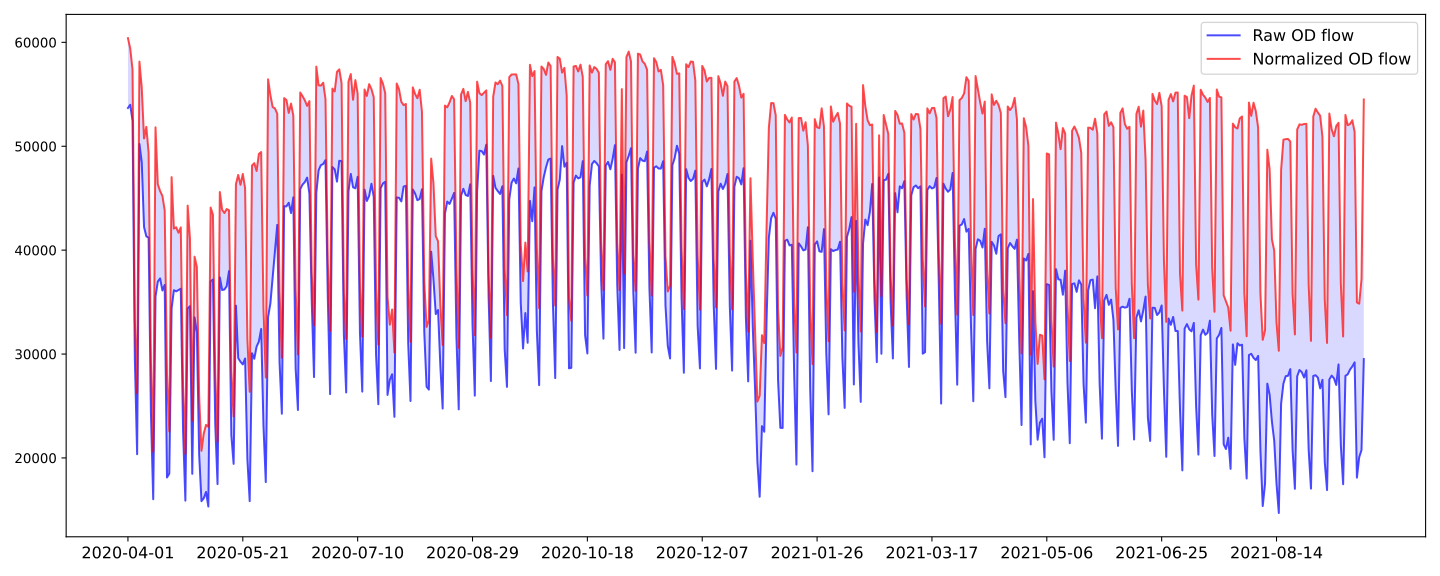}
    \caption{Dynamic flow from Saitama to Tokyo from 2020/04/01 to 2021/09/21 before normalization (blue) and after normalization (red).}
    \label{fig:normalization}
\end{figure}

\section{Data}\label{sec:data}
We collect multi-source epidemic-related data and mobility data for epidemic prediction. We set 47 prefectures of Japan and 2020/04/01 $\sim$ 2021/09/21 (539 days) as our study area and time period, respectively. The used data can be divided into three sub-groups, including epidemic data, external data and mobility flow data. We will describe the data sources, demonstrate the data processing and visualize the data in this section.

\subsection{Epidemic Data}\label{sec:data-epidemic_data}
The number of daily confirmed cases and cumulative cases and deaths are collected from the NHK COVID-19 database\footnote{https://www3.nhk.or.jp/news/special/coronavirus/data/}. Fig. \ref{fig:spatial_distribution} demonstrates the spatial distribution of daily confirmed cases of Japan on 2021/04/01 and 2021/08/01, respectively. 
Fig. \ref{fig:tokyo_distribution} demonstrates the daily confirmed cases of Tokyo and the distribution difference between two periods of time (2020/04/01 $\sim$ 2021/06/30 and 2021/07/01 $\sim$ 2021/09/21).

The number of recovered cases is collected from Japan LIVE Dashboard\footnote{https://github.com/swsoyee/2019-ncov-japan} \cite{ref_25} (original data source is from Ministry of Health, Labour and Welfare, Japan). The population of each prefecture is collected from 2020 Japan census data\footnote{https://www.stat.go.jp/data/kokusei/2020/kekka.html}. With above-mentioned data, daily $S$, $I$, $R$ of each prefecture can be calculated.

\subsection{External Data}\label{sec:data-external data}
Apart from the number of daily confirmed cases, the input node features also include daily movement change, the ratio of daily confirmed cases in active cases, and $dayofweek$. The ratio of daily confirmed cases in active cases can reflect the trend of epidemic which is very important for epidemic prediction, especially for periods near change points.

The movement change data is collected from Facebook Movement Range Maps\footnote{https://data.humdata.org/dataset/movement-range-maps}. It records the change of people movement range compared to a baseline period. Because it is not provided at prefecture level, we use population weighted average to get data at prefecture level. Fig. \ref{fig:movement_popweight} visualizes the original movement change data and the prefecture-level movement change data through population weighted averaging on 2021/08/01. Because there exists the weekly periodicity in number of people taken test, it is necessary to include $dayofweek$ as one of external features.

\subsection{Mobility Flow Data}\label{sec:data-mobility_flow_data}
The input static flow data for adaptive graph learning module is the number of commuters between each pair of prefectures, which is collected from 2015 Japan census data\footnote{https://www.stat.go.jp/data/kokusei/2015/kekka.html}. 

The input dynamic flow data for dynamic graph learning module is the daily OD flow data among 47 prefectures, which is generated from human GPS trajectory data provided by Blogwatcher Inc.. Blogwatcher dataset provides the GPS trajectories of multi-million users containing the rich human mobility information of Japan. We use detected Move/Stay points to divide each trajectory into several OD trips and aggregate the trips by origin prefecture and destination prefecture of each trip to get the daily dynamic OD flow between each pair of prefectures. 

However, the significant spatio-temporal imbalance exists in daily dynamic OD flow data:

(1) Since the GPS data is collected from mobile application users, there are significant gaps among the utilization rates (i.e., sample rate of GPS data) of different prefectures;

(2) The number of unique user IDs in GPS records keeps varying every day, and the trend and periodicity exist in the variations. 

It would be problematic to use the raw dynamic OD flow data directly which could lead to incorrect representation of human mobility intensity between regions and mislead the construction of epidemic propagation graph. We need to normalize the raw dynamic flow data to mitigate the spatio-temporal imbalance. However, it is extremely difficult to deal with spatial imbalance and temporal imbalance simultaneously. Because temporal imbalance itself is also spatially different, and vice versa. 

Normalizing the raw dynamic flow data only using GPS data would be very complicated. We address this problem by introducing the extra data, stay put ratio (ratio of people staying in a single location all day) data in Facebook Movement Range Maps. As mentioned in Sec. \ref{sec:data-external data}, we also use population weighted average to get the stay put ratio on prefecture level. Then, a simple method can be applied to normalize the dynamic flow data:

(1) Using stay put ratio and population, we can get the proportion of active people (people not staying in a single location all day) by:
\begin{equation}
    \begin{aligned}
        \widetilde{P}^{t}_{n} = (1 - {stay}^{t}_{n}) \times P_n 
    \end{aligned}
\end{equation}
where $\widetilde{P}^{t}_{n}$ is the number of active people of region $n$ at time $t$ and ${stay}^{t}_{n}$ is stay put ratio of region $n$ at time $t$.

(2) We aggregate the trips by their origin regions, and count the number of unique user IDs in each group of trips with same origin region. Then, the sample rate of dynamic flow data can be computed by: 
\begin{equation}
    \begin{aligned}
        {sample}^{t}_{n} = \frac{{nuid}_{n,:}^t}{\widetilde{P}^{t}_{n}}
    \end{aligned}
\end{equation}
where ${sample}^{t}_{n}$ is the sample rate of region $n$ at time $t$ and ${{nuid}_{n,:}^t}$ is the number of unique user IDs in the group of trips with same origin $n$ at time $t$.

(3) With the sample rate of each region, we can set an anchor using sample rate of a specific region at a specific timestep (in our case, the sample rate of Tokyo at 2020/04/01) and normalize the raw dynamic OD flow data by:
\begin{equation}
    \begin{aligned}
        \mathcal{O}_{n,m}^t = \frac{{sample}^{t_a}_{n_a}}{{sample}^{t}_{n}} \times \overline{\mathcal{O}}_{n,m}^t
    \end{aligned}
\end{equation}
where ${sample}^{t_a}_{n_a}$ is the anchor (the sample rate of region $n_a$ at time $t_a$), $\overline{\mathcal{O}}_{n,m}^t$ is the raw dynamic OD flow from region $n$ to $m$ at time $t$, and $\mathcal{O}_{n,m}^t$ is normalized dynamic flow data which can form the input to dynamic graph learning module $\mathcal{O}^{t-(T_{in}-1):t} \in \mathbb{R}^{N\times N \times T_{in}}$.

Although this method may not perfectly remove the spatio-temporal imbalance in dynamic OD flow data, it is a simple but effective way to normalize the dynamic OD flow data. We demonstrate its effectiveness in Fig. \ref{fig:normalization}.

By combining the above-mentioned epidemic data, external data and mobility flow data, finally we can get all the data for experiments. In cases of prefecture-level epidemic forecasting task of Japan, the input features of 47 prefectures are generated as a (539, 47, 4) tensor, the static flow is a (47, 47) matrix, and the dynamic flow is a (539, 47, 47) tensor. 

\section{Experiments}\label{sec:experiment}
In this study, we use the collected and processed epidemic data, external data and mobility flow data from 2020/04/01 to 2021/09/21 for epidemic forecasting tasks of 47 prefectures in Japan. We conduct the experiments on this epidemic prediction tasks by comparing three classes of baseline models (including mechanistic models, general spatio-temporal deep learning models and GNN-based epidemic models). Furthermore, we use case studies to demonstrate and analyze the high reliability and interpretability of our model. In addition, as the initialization of adaptive graph learning in our model with minimal data requirement, the effectiveness of generated mobility data by the mobility generation method is also evaluated by experiments.

\begin{table*}[ht]  
	\centering  
	\fontsize{10}{12}\selectfont 
		\caption{Performance Comparison with Baselines}  
		\label{tab:performance_comparison}  
		\resizebox{\textwidth}{!}{
		\begin{tabular}{ccccccccc}  
			\toprule  
            \multirow{2}{*}&  
			\multicolumn{4}{c}{3 Days Ahead}&\multicolumn{4}{c}{ 7 Days Ahead}\cr  
			\cmidrule(lr){2-5} \cmidrule(lr){6-9}  
			Model&RMSE&MAE&MAPE&RAE&RMSE&MAE&MAPE&RAE\cr 
			\midrule  
			SIR&429.4±23.2&153.9±5.2&83.8±0.7&0.47±0.02&507.5±29.6&191.4±7.7&111.4±3.8&0.57±0.02\cr
			SIR(Copy)&248.1&97.4&57.4&0.29&318.5&127.1&67.2&0.38\cr
			MetaSIR&336.0±21.6&126.8±3.5&72.2±0.9&0.38±0.01&429.8±25.5&166.9±3.7&92.9±0.8&0.50±0.01\cr 
			MetaSIR(Copy)&236.5&92.2&54.1&0.28&307.6&120.0&62.7&0.36\cr \cmidrule(lr){1-9}
			STGCN&375.6±18.8&118.6±10.8&45.3±2.8&0.36±0.03&381.1±17.7&128.0±6.6&52.5±3.0&0.38±0.02\cr
			DCRNN&305.0±9.8&89.3±4.4&37.3±0.7&0.27±0.01&323.8±15.9&107.6±5.3&47.3±1.4&0.32±0.02\cr
			AGCRN&223.5±28.5&80.0±7.8&56.6±13.2&0.24±0.02&253.1±37.7&97.9±7.6&60.8±10.1&0.29±0.02\cr
			GraphWaveNet&223.8±46.6&70.6±11.7&35.4±1.2&0.21±0.04&259.9±52.2&89.2±15.2&42.3±1.5&0.27±0.05\cr
			MTGNN&297.6±19.2&102.4±6.7&40.6±0.8&0.31±0.02&363.5±37.9&130.9±13.1&49.1±1.7&0.39±0.04\cr \cmidrule(lr){1-9}
			CovidGNN&261.9±55.5&88.4±16.7&43.3±3.8&0.27±0.05&305.4±70.6&116.5±23.8&60.9±5.3&0.35±0.07\cr
			ColaGNN&221.7±40.7&72.7±7.2&38.9±1.5&0.22±0.02&300.6±61.2&109.4±16.4&49.3±1.5&0.33±0.05\cr \cmidrule(lr){1-9}
			MepoGNN(Adp)&\underline{141.0±7.2}&\underline{54.3±2.3}&\underline{34.9±0.8}&\underline{0.16±0.01}&\underline{174.6±10.1}&\underline{69.7±4.2}&\textbf{41.4±1.6}&\underline{0.21±0.01}\cr
			MepoGNN(Dyn)&\textbf{135.9±17.8}&\textbf{52.7±4.6}&\textbf{34.2±0.7}&\textbf{0.16±0.01}&\textbf{160.6±4.5}&\textbf{67.6±1.2}&\underline{41.7±0.9}&\textbf{0.20±0.00}\cr
			\toprule
			\multirow{2}{*}&  
			\multicolumn{4}{c}{ 14 Days Ahead}&\multicolumn{4}{c}{  Overall}\cr  
			\cmidrule(lr){2-5} \cmidrule(lr){6-9}  
			Model&RMSE&MAE&MAPE&RAE&RMSE&MAE&MAPE&RAE\cr 
			\midrule  
			SIR&890.2±83.8&314.5±16.9&228.3±11.8&0.94±0.05&595.0±43.5&210.0±9.2&128.2±4.7&0.63±0.03\cr			SIR(Copy)&835.5&332.6&183.2&1.00&539.1&190.2&102.7&0.57\cr
			MetaSIR&766.1±58.5&279.1±8.2&177.4±4.5&0.84±0.02&500.4±33.9&182.1±4.4&104.9±1.3&0.55±0.01\cr 			MetaSIR(Copy)&786.4&302.7&161.9&0.91&503.7&175.6&92.7&0.53\cr \cmidrule(lr){1-9}
			STGCN&430.2±15.8&159.4±6.0&74.7±3.7&0.48±0.02&389.5±7.9&132.0±2.9&55.6±2.4&0.40±0.01\cr
			DCRNN&377.9±11.1&146.0±5.0&69.5±4.0&0.44±0.01&335.0±11.8&112.5±4.5&49.5±1.3&0.34±0.01\cr
			AGCRN&390.4±105.8&149.0±11.4&88.0±12.8&0.45±0.03&322.7±136.7&108.0±9.9&67.9±15.6&0.32±0.03\cr
			GraphWaveNet&389.8±20.8&144.4±7.3&\underline{60.2±4.2}&0.43±0.02&294.7±40.9&100.1±11.1&44.7±1.4&0.30±0.03\cr
			MTGNN&443.5±15.4&168.3±8.1&68.0±2.9&0.50±0.02&363.2±20.5&130.0±8.3&50.7±1.6&0.39±0.03\cr \cmidrule(lr){1-9}
			CovidGNN&414.7±59.8&177.4±15.9&111.2±6.6&0.53±0.05&329.6±59.8&124.2±19.2&66.9±4.2&0.37±0.06\cr
			ColaGNN&388.3±23.2&153.4±10.2&75.5±10.8&0.46±0.03&310.7±31.4&110.2±7.2&51.9±3.7&0.33±0.02\cr \cmidrule(lr){1-9}
			MepoGNN(Adp)&\underline{261.1±16.0}&\textbf{105.1±7.3}&\textbf{60.1±3.2}&\textbf{0.32±0.02}&\underline{196.2±11.3}&\underline{75.4±4.7}&\textbf{44.0±1.6}&\underline{0.23±0.01}\cr
			MepoGNN(Dyn)&\textbf{253.2±7.5}&\underline{107.0±3.0}&62.0±2.0&\underline{0.32±0.01}&\textbf{186.1±5.0}&\textbf{74.3±2.0}&\underline{44.4±0.8}&\textbf{0.22±0.01}\cr
			\bottomrule 
		\end{tabular}}
\end{table*}

\subsection{Setting}\label{subsec:setting}
The input time length $T_{in}$ and output time length $T_{out}$ are both set to 14 days which means we use two-week historical observations to do the two-week prediction of daily confirmed cases. Then, we split the data with ratio 6:1:1 to get training/validation/test datasets, respectively. The fifth wave of infection in Japan is included in test dataset to test the model performance on a real outbreak situation. During training, we use the curriculum learning strategy \cite{ref_26} which increases one prediction horizon every two epochs starting from one day ahead prediction until reaching output time length. The batch size is set to 32. The loss function is set as \textit{MAE} (Mean Absolute Error). Adam is set as the optimizer, where the learning rate is 1e-3 and weight decay is 1e-8. The training algorithm would either be early-stopped if the validation error did not decrease within 20 epochs or be stopped after 300 epochs. PyTorch is used to implement our model. Then experiments are performed on a server with four 2080Ti GPUs. 

Finally, we evaluate the performance of model on 3 days, 7 days, 14 days ahead prediction and overall 14 steps prediction. The four metrics are used to qualify the performance: \textit{RMSE} (Root Mean Square Error; as Eq. \ref{RMSE}), \textit{MAE} (Mean Absolute Error; as Eq. \ref{MAE}), \textit{MAPE} (Mean Absolute Percentage Error; as Eq. \ref{MAPE}) and \textit{RAE} (Relative Absolute Error; as Eq. \ref{RAE}). To mitigate the influence of randomness, we perform 5 trials for each model and calculate the mean and 95\% confidence interval of results. The used random seeds are 0, 1, 2, 3, 4. 
\begin{equation}\label{RMSE}
    RMSE(\hat{\mathbf Y},\mathbf Y) = \sqrt{\frac{1}{NT_{out}}\sum_{i=1}^{N}\sum_{j=1}^{T_{out}} (\hat{\mathbf Y}_i^{t+j} - \mathbf Y_i^{t+j})^2}
\end{equation}
\begin{equation}\label{MAE}
    MAE(\hat{\mathbf Y},\mathbf Y) = \frac{1}{NT_{out}}\sum_{i=1}^{N}\sum_{j=1}^{T_{out}}\lvert \hat{\mathbf Y}_i^{t+j} - \mathbf Y_i^{t+j} \rvert
\end{equation}
\begin{equation}\label{MAPE}
    MAPE(\hat{\mathbf Y},\mathbf Y) = \frac{100\%}{NT_{out}}\sum_{i=1}^{N}\sum_{j=1}^{T_{out}}\Bigl| \frac{\hat{\mathbf Y}_i^{t+j} - \mathbf Y_i^{t+j}}{\mathbf Y_i^{t+j}}\Bigr|
\end{equation}
\begin{equation}\label{RAE}
    RAE(\hat{\mathbf Y},\mathbf Y) = \frac{\sum_{i=1}^{N}\sum_{j=1}^{T_{out}}\lvert \hat{\mathbf Y}_i^{t+j} - \mathbf Y_i^{t+j}\rvert}  {{\sum_{i=1}^{N}\sum_{j=1}^{T_{out}}\lvert \mathbf Y_i^{t+j} - \overline{\mathbf Y_{1:N}^{t+1:t+T_{out}}}}\rvert} 
\end{equation}

\subsection{Evaluation}
We implement three classes of baseline models (including mechanistic models, general spatio-temporal deep learning models and GNN-based epidemic models), compare them with our model, and evaluate the performance on epidemic prediction task of 47 prefecture in Japan.

\begin{itemize}
    \item \noindent\textbf{Mechanistic Models:}
    
    \textbf{(1) SIR} \cite{ref_11}. SIR model is one of most basic compartmental models in epidemiology. We use optimized $\beta$ and $\gamma$ of each region to produce the prediction.

    \textbf{(2) SIR(Copy)}. Because of weekly periodicity, we copy the $\beta$ and $\gamma$ of last week to produce the prediction. 
    
    \textbf{(3) MetaSIR} \cite{ref_2}. Metapopulation SIR model considers the heterogeneity of sub-populations and models the interaction between sub-populations. We use the commuter survey data as $\mathcal H$ and optimize $\beta$ and $\gamma$ for each region to produce the prediction. 
    
    \textbf{(4) MetaSIR(Copy)}. We copy the $\beta$ and $\gamma$ of last week to produce the prediction. 

    \item\noindent\textbf{Spatio-temporal Deep Learning Models:} 
    
    \textbf{(5) STGCN} \cite{ref_21}. STGCN is one of the earliest models which applies GCN and TCN to do spatio-temporal prediction. 
    
    \textbf{(6) DCRNN} \cite{ref_22}. DCRNN proposes a variant of GCN, called diffusion convolution and combines it with gated recurrent unit (GRU) to build a spatio-temporal prediction model. 
    
    \textbf{(7) GraphWaveNet} \cite{ref_23}. GraphWaveNet proposes an adaptive learnable graph and uses GCN and TCN to capture spatio-temporal dependency.  
    
    \textbf{(8) MTGNN} \cite{ref_26}. MTGNN uses a graph learning module to learn spatial correlation and fuse different spatial hops and different TCN kernels to enhance the model capacity. 
    
    \textbf{(9) AGCRN} \cite{ref_27}. AGCRN uses GCN and GRU along with a graph learning module and a node adaptive parameter learning module to capture spatio-temporal dependency.

    \item\noindent\textbf{GNN-based Epidemic Models:} 
    
    \textbf{(10) CovidGNN} \cite{ref_10}. CovidGNN is one of the earliest GNN-based epidemic models. It embeds temporal features on each node and uses GCN with skip connections to capture spatial dependency. 
    
    \textbf{(11) ColaGNN} \cite{ref_9}. ColaGNN uses the location-aware attention to extract spatial dependency and uses GCN to integrate the spatio-temporal information.

\end{itemize}

\begin{figure}[t]
    \centering
    \includegraphics[width=1.0\linewidth]{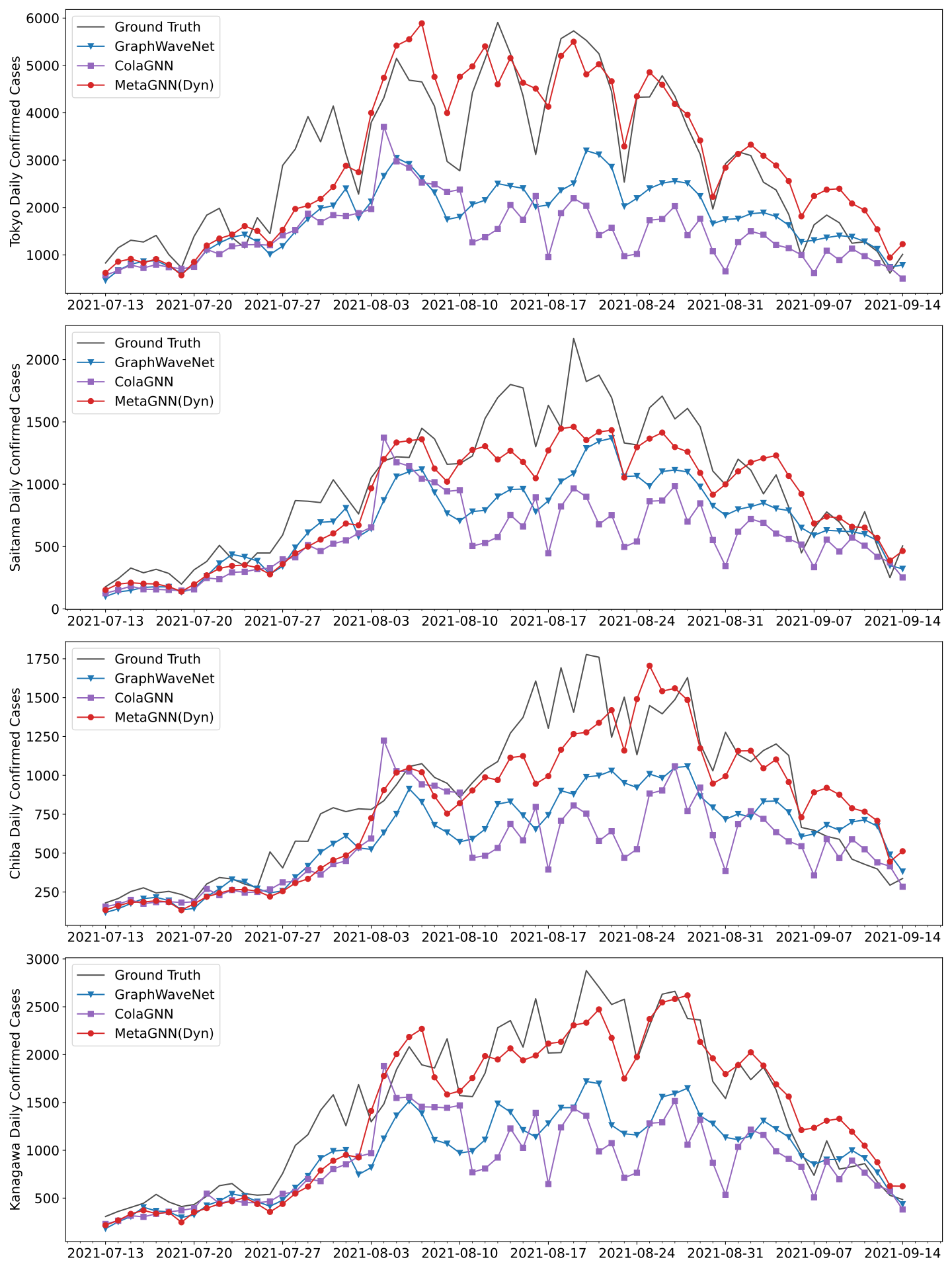}
    \caption{Predicted daily confirmed cases of Tokyo, Saitama, Chiba and Kanagawa with horizon=7.}
    \label{fig:Tokyo_Saitama_Chiba_Kanagawa}
\end{figure}

\noindent\textbf{Performance Evaluation.} In Table \ref{tab:performance_comparison}, we compare the performance on three different horizons and overall performance for multi-step prediction among the above-mentioned three classes of baseline models and proposed MepoGNN with two types of graph learning module. Generally, the spatio-temporal deep learning models and GNN-based epidemic models outperform the mechanistic models, especially for longer horizons. Among all baseline models, GraphWaveNet gets the best performance. However, our proposed two MepoGNN models get the very significant improvement over all baseline models. For two types of graph learning module, the dynamic one gets slightly better performance than the adaptive one. Fig. \ref{fig:Tokyo_Saitama_Chiba_Kanagawa} compares the 7 days ahead prediction results of Tokyo, Saitama, Chiba and Kanagawa of the top two baseline models and MepoGNN model with dynamic graph learning module. From the prediction results, GraphWaveNet and ColaGNN can not produce accurate predictions for high daily confirmed cases during the epidemic outbreak. This phenomenon could be explained by different data distributions of daily confirmed cases in training dataset and test dataset. The test dataset covers the period of fifth epidemic wave in Japan which is much more severe than previous ones. Deep learning models have difficulty to predict these high daily confirmed cases that never happened before the fifth wave. However, with the help of metapopulation SIR module, our proposed MepoGNN model can handle this problem and make significantly better prediction for unprecedented surge of cases. This capability is very crucial for a trustworthy epidemic forecasting model. 

\begin{table*}[t]  
	\centering
	\fontsize{10}{12}\selectfont 
		\caption{Ablation Study}  
		\label{tab:Ablation_study}  
        \begin{tabular}{cccccc}
            \toprule
           Graph&Model&Mean RMSE&Mean MAE&Mean MAPE&Mean RAE \\ \midrule
        \multirow{4}{*}{Adaptive}&w/o glm&209.51±22.70&81.85±6.69&47.51±2.62&0.25±0.02\\
                    &w/o propagation&203.23±24.70&82.05±8.05&45.84±1.68&0.25±0.02\\
                    &w/o SIR&318.05±16.30&108.53±5.26&46.07±0.53&0.33±0.02\\               &MepoGNN&\textbf{196.16±11.33}&\textbf{75.45±4.65}&\textbf{44.02±1.55}&\textbf{0.23±0.01}\\\cmidrule(lr){1-6} 
            \multirow{4}{*}{Dynamic}&w/o glm&194.50±17.65&76.84±6.04&\textbf{43.63±1.59}&0.23±0.02\\
                    &w/o propagation&200.55±17.00&80.73±5.54&45.16±1.24&0.24±0.01\\
                    &w/o SIR&290.78±33.92&102.00±9.93&45.79±1.61&0.31±0.03\\
                    &MepoGNN&\textbf{186.07±4.99}&\textbf{74.30±1.99}&44.43±0.77&\textbf{0.22±0.01}\\
                    \bottomrule
\end{tabular} 
\end{table*}

\noindent\textbf{Ablation Study.} To demonstrate the effectiveness of different components of our model, we conduct an ablation study for MepoGNN models with two different graph learning modules, respectively. The variants are as follows: 

\textbf{(1) w/o glm}: Remove the graph learning module of MepoGNN model; 

\textbf{(2) w/o propagation}: Remove the metapopulation propagation from metapopulaiton SIR module (which means metapopulation SIR model is reduced to SIR model); 

\textbf{(3) w/o SIR}: Remove the metapopulation SIR module completely. 

\begin{figure}[t]
    \centering
    \includegraphics[width=1.0\linewidth]{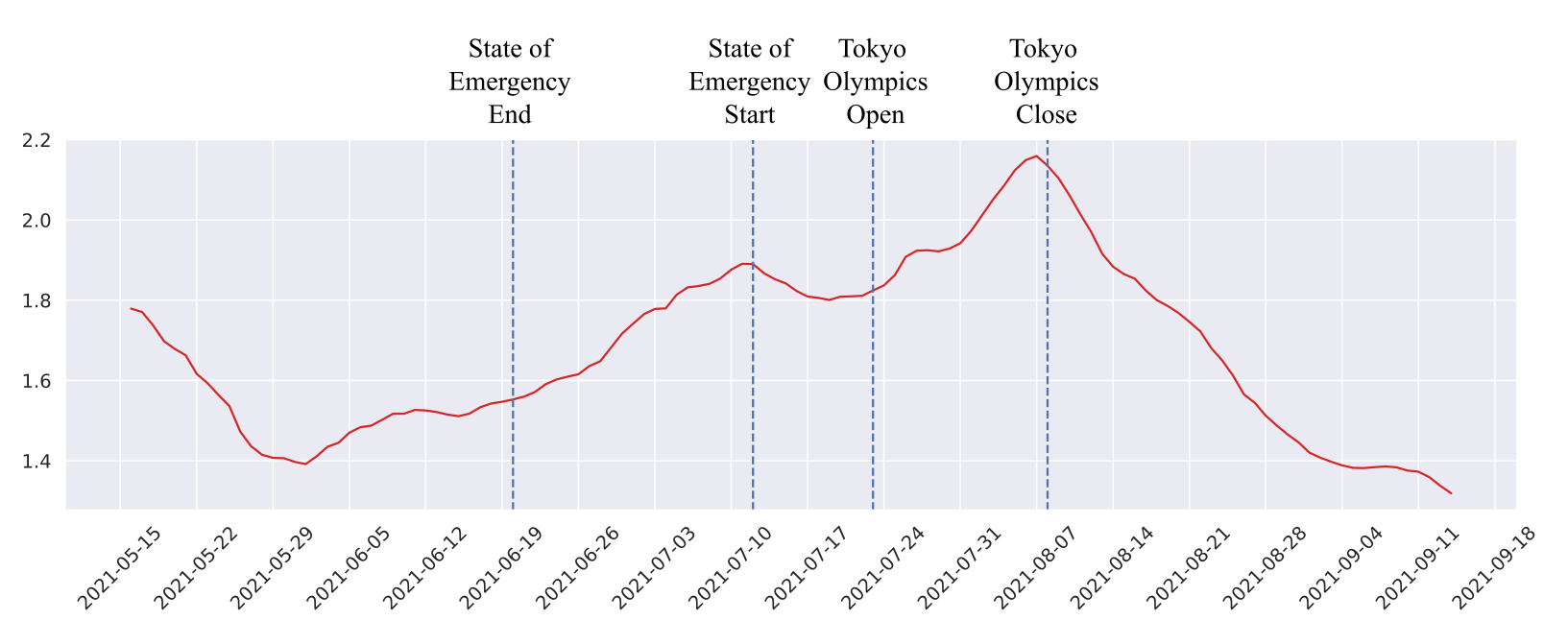}
    \caption{7-day moving average of predicted $\hat{R}^t$ of Tokyo with horizon=7.}
    \label{fig:beta}
\end{figure}
\begin{figure}[t]
    \centering
    \includegraphics[width=0.6\linewidth]{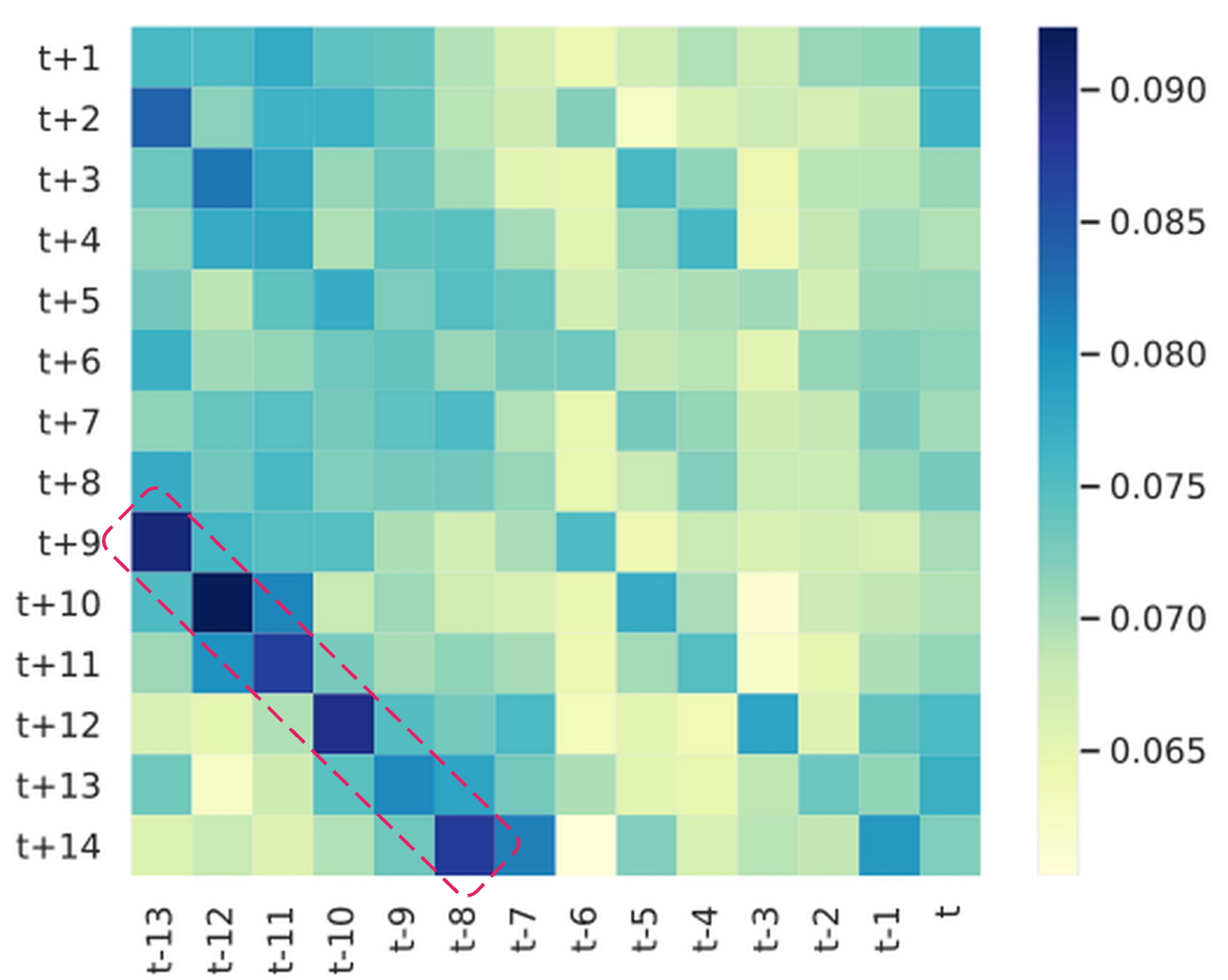}
    \caption{Learned time weight matrix in dynamic graph learning module.}
    \label{fig:incidencegraph}
\end{figure}

Table \ref{tab:Ablation_study} demonstrates that all three components can bring significant boost of performance to our model. Particularly, it is easy to find that the biggest performance drop happens when removing the metapopulation SIR module. Because the metapopulation SIR module enables the capability of MepoGNN model to handle the unprecedented surge of cases.

\begin{figure}[t]
    \centering
    \includegraphics[width=1.0\linewidth]{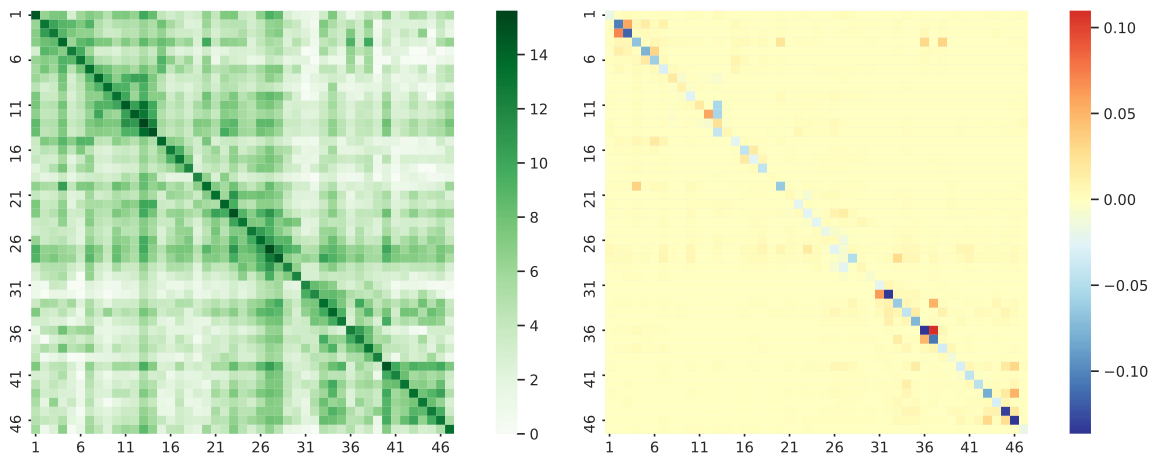}
        \caption{Learned adaptive mobility graph of the 47 prefectures of Japan with log transformation (left) and its difference with static commuter graph (right).}
    \label{fig:adaptivegraph}
\end{figure}

\subsection{Case Study}
The final output of MepoGNN model is fully produced by metapopulation SIR module. It brings significant interpretability to our model. We conduct an analysis for the predicted parameters of metapopulation SIR module to demonstrate the interpretability.

As shown in Fig. \ref{fig:beta}, we plot weekly average of predicted pseudo effective reproduction number\footnote{$\beta$ in metapopulation SIR model is not completely equivalent to $\beta$ in SIR model, so we call $\hat{R}^t =\beta^t/\gamma^t$ pseudo effective  reproduction number.} ($\hat{R}^t =\beta^t/\gamma^t$) of Tokyo at 7 days ahead horizon in validation and test dataset and label major events and policy changes on timeline. $\hat{R}^t$ starts to increase when state of emergency ends and starts to decrease when state of emergency starts. $\hat{R}^t$ rapidly increases during Tokyo Olympics, and decreases after it. It demonstrates the change of predicted $\hat{R}^t$ is consistent with real events and policy changes.  

Fig. \ref{fig:incidencegraph} shows the learned time weight matrix of dynamic graph learning module. The most significant time lag of mobility effect on epidemic spread is 22 days. This result is consistent with a public health research \cite{ref_28} which states that the effective reproduction number significantly increased 3 weeks after the nightlife places mobility increased in Tokyo. Although the used indicator is different from our research, the mechanisms behind the time lag could be similar. 

Fig. \ref{fig:adaptivegraph} shows the learned graph of adaptive graph learning module and the differences between it and the commuter graph used as initialization. The learned adaptive mobility graph keeps the major structure of commuter graph. And the minor changes from initialization can reflect the pattern differences between regular commuting and spatial epidemic propagation.

\subsection{Mobility Generation Test}
We introduce two different types of graph learning module (Dynamic and Adaptive) in our MepoGNN model to handle different data accessibility of flow data. Furthermore, in the experiments of epidemic prediction for 47 prefectures of Japan, we collect the static flow data and dynamic flow data as inputs to adaptive and dynamic graph learning module, respectively. However, even static flow data which is used to initialize the learnable graph of adaptive graph learning module is not always accessible in real world. So, we conduct an additional experiment to evaluate the performance of our MepoGNN model under the the situation of worse data availability. Without using any mobility flow data, we just use population of each prefecture and distance between each pair of prefectures to generate the mobility graph for adaptive graph learning module and then compare the performance of initialization from static flow data and generated mobility data. 

By using Eq. \ref{relativemob}, we can generate the relative mobility intensity between each pair of prefectures. In this experiment, we set parameter $\alpha$ as 1e-6, $d$ as 1.7 and $\epsilon$ as 9. Then, the relative mobility intensity between prefecture $n$ and $m$ can be computed by: 
\begin{equation}
    \widetilde{mob}_{nm} = 10^{-6} \times \frac{P_n P_m}{(dist_{nm})^{1.7} + 9}
\end{equation}

Then, we use the relative mobility intensity matrix $\mathbf{\widetilde M} \in \mathbb{R}^{N \times N}$ to initialize the learnable graph $\mathbf{G}$ in adaptive graph learning module and run the experiments using same setting mentioned in Section \ref{subsec:setting}. We compare the performance of two types of initialization (using static flow matrix or generated relative mobility intensity matrix) for adaptive graph learning module as Table \ref{tab:static_generate}. The performance of initialization using mobility generation is competitive, compared with that of initialization using static flow data. Since the performance gap is very small, the mobility generation method can be considered as a reliable alternative to static flow data serving as the input to adaptive graph learning module. Furthermore, with minimal data requirement, this mobility generation method can help us handle the situation of lacking adequate data (i.e., flow data) and make our model much more applicable. 

\begin{table*}[t!]  
	\centering  
	\fontsize{10}{12}\selectfont 
		\caption{Adaptive Graph Initialization Comparison}  
		\label{tab:static_generate}  
        \begin{tabular}{cccccc}
            \toprule
           Model&Initialization&Mean RMSE&Mean MAE&Mean MAPE&Mean RAE \\ \midrule
            MepoGNN(Adp)&Static Flow Matrix&196.16±11.33&75.45±4.65&44.02±1.55&0.23±0.01 \\
                    &Mobility Generation&195.56±6.55&75.84±2.84&47.77±2.65&0.23±0.01\\\cmidrule(lr){1-6}
\end{tabular} 
\end{table*}

\begin{figure}[t]
    \centering
    \includegraphics[width=1\linewidth]{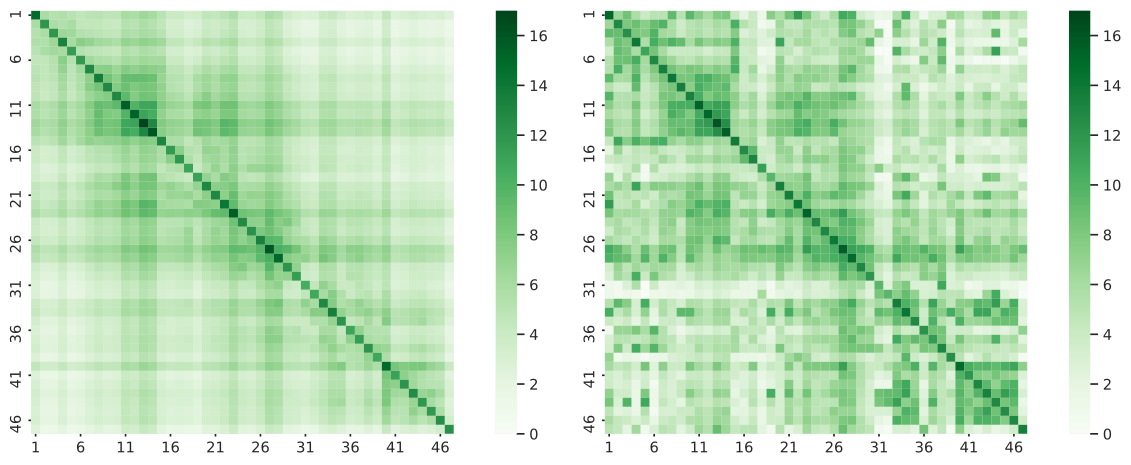}
        \caption{Generated mobility matrix (left) and learned adaptive mobility graph using it as initialization (right) with log transformation.}
    \label{fig:generate}
\end{figure}
Additionally, we also visualize the generated mobility matrix and learned graph using initialization of generate mobility matrix as Fig. \ref{fig:generate}. The generated mobility matrix is symmetric and relatively smooth as initialization, but the learned graph shows directional and complex spatial structure. This phenomenon demonstrates that the complex spatial correlation of epidemic propagation can be learned during training in our model, even with the initialization by a mobility generation method with minimal data requirement.

\begin{figure}[t]
    \includegraphics[width=1.0\linewidth]{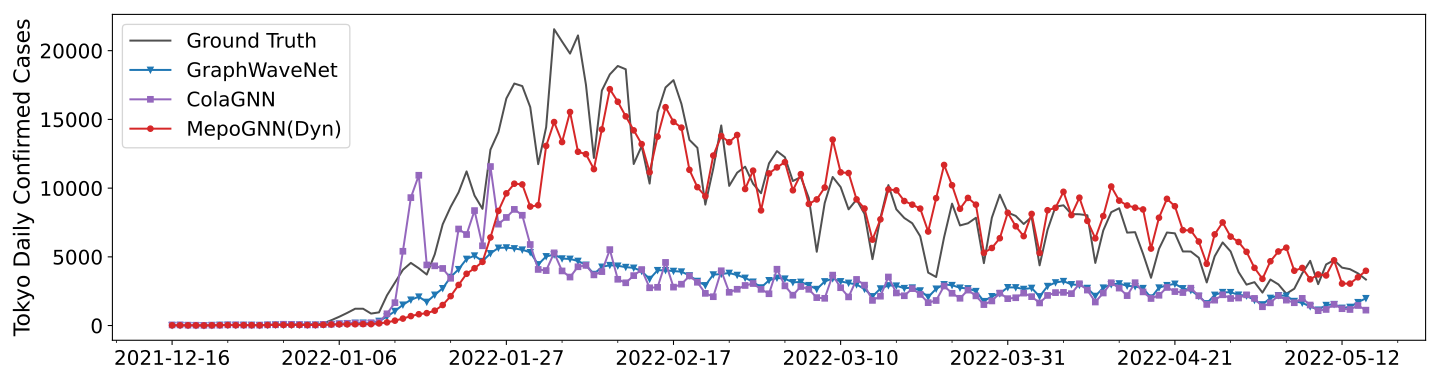}
    \caption{Predicted daily confirmed cases of Tokyo with horizon=7 on the extra test dataset from the sixth epidemic wave.}
    \label{fig:newdata}
\end{figure}

\section{Conclusion}
Since the outbreak of COVID-19, epidemic forecasting has become a key research topic again. In this study, we propose a novel hybrid model called MepoGNN for epidemic forecasting that incorporates spatio-temporal graph neural networks and graph learning mechanisms into metapopulation SIR model. To the best of our knowledge, our model is the first hybrid model that couples metapopulation epidemic model with spatio-temporal graph neural networks. Our model can not only predict the number of confirmed cases but also explicitly learn the time/region-varying epidemiological parameters and the underlying epidemic propagation graph from heterogeneous data in an end-to-end manner. 

Then, we collect and process the real data under COVID-19 from 2020/04/01 to 2021/09/21 in Japan, including epidemic data, external data and mobility flow data, for epidemic forecasting task. We evaluate our model by comparing with three classes of baseline models (including mechanistic models, GNN-based epidemic models and general spatio-temporal deep learning models) on this task. The results demonstrate that our model outperforms all baseline models and have the capability to handle the unprecedented surge of cases. We also visualize the learned parameter and epidemic propagation graph in case studies to illustrate the high interpretability of our model. Besides building and evaluating the proposed MepoGNN model, we additionally propose a mobility generation method with minimal data requirement to deal with the situations that the mobility data is unavailable. And the experimental results demonstrate the effectiveness of using generated mobility in our model. 

This work demonstrates combining deep learning model and domain knowledge can bring great benefits to the performance, reliability and interpretability of model. So, a potential future direction could be exploring this kind of combination in other fields (e.g., incorporating the domain knowledge of traffic into spatio-temporal traffic prediction models).

\section{Limitation}
One of the limitations is that our proposed method can not perfectly handle the highly extreme situations of sudden outbreak. Fig. \ref{fig:newdata} shows the predictions in the sixth epidemic wave of Tokyo in which the number of confirmed cases suddenly surged to thousands from continued near zero in a very short period of time. The proposed method can outperform baselines in this case, but it fails to produce accurate predictions at the beginning of this epidemic wave. Although difficult, collecting extra data and improving the model to predict the extreme outbreak would be an important research direction worth further investigations.

\section*{Acknowledgments}
This work was partially supported by JST SICORP Grant Number JPMJSC2104. 

Part of this work first published in [Vol. 13718, pp. 453-468, 2023] by Springer Nature \cite{cao2023mepognn}.

\bibliographystyle{IEEEtran}
\bibliography{references}

\vfill

\end{document}